\documentclass[%
aip,
amsmath,amssymb,
reprint,%
]{revtex4-1}
\usepackage{epsfig, subfigure}
\usepackage{verbatim}
\usepackage{color}
\usepackage{amsmath}
\usepackage{booktabs}
\usepackage{threeparttable}
\usepackage{amsthm}
\graphicspath{{figs/}}

\newtheorem{mypro}{Proposition}

\begin{document}
\title{Mechanism of separation hysteresis in curved compression ramp}

\author{Wen-Feng Zhou}
\author{Yan-Chao Hu}
\email[]{huyanchao@cardc.cn}
\author{Ming-Zhi Tang\textsuperscript{1}, Gang Wang\textsuperscript{1}, Ming Fang} 
\affiliation{Hypervelocity Aerodynamics Institute, China Aerodynamics Research and Development Centre, Mianyang 621000, China}
\author{Yan-Guang Yang}
\email[]{yanguangyang@cardc.cn}

\affiliation{China Aerodynamics Research and Development Centre, Mianyang 621000, China}


\date{\today}

\begin{abstract}
A new spatial-related mechanism is proposed to understand separation hysteresis processes in curved compression ramp (CCR) flows discovered recently (Hu et al. Phy. Fluid, 32(11): 113601, 2020). Two separation hystereses, induced by variations of Mach number $M_{\infty}$ and wall temperature $T_w$, are investigated numerically. The two hystereses indicate that there must exist parameter intervals of Mach number $[M_{\infty,s},M_{\infty,a}]$ and wall temperature $[T_{w,a},T_{w,s}]$, wherein both attachment and separation states can be established stably. The relationships between the aerodynamic characteristics (including wall friction, pressure and heat flux) and the shock wave configurations in this two hystereses are analyzed. Further, the adverse pressure gradient (APG) $I_{sb}(x)$ induced by the upstream separation process and APG $I_{cw}(x)$ induced by the downstream isentropic compression process are estimated by classic theories. The trend of boundary layer APG resistence $I_{b}(x)$ is evaluated from the spatial distributions of the physical quantities such as the shape factor and the height of the sound velocity line. With the stable conditions of separation and attachment, a self-consistent mechanism is obtained when $I_{sb}$, $I_{cw}$ and $I_{b}$ have appropriate spatial distributions.



\end{abstract}


\maketitle


\section{Introduction}
Wide speed range and long endurance are two key requirements in future flight\cite{dolling2001fifty,gui2019combined,zhen-guo2010key}. The corresponding large variations of inflow Mach number $Ma_{\infty}$ and normalized wall temperature $T_{w}$ lead to multiple complex effects on the aerodynamic force and heat, e.g., a great decreasing of inlet's capture flow ability, generation of a large additional drag\cite{babinsky2011shock,delery1985shock,green1970interactions,zhang2020hypersonic}. Hysteresis is one of the intrinsic cause, which is a general property of multistable systems manifested as the dependence of a system on its evolutionary history, and ubiquitous in aerospace flow systems\cite{hu2020bistable}. It is found that the aerodynamic coefficients have multiple values in angle of attack hysteresis loop\cite{yang2008experimental,mccroskey1982unsteady,mueller1985influence,biber1993hysteresis,mittal2000prediction}. For the shock wave reflection, regular and Mach reflection hystersis is familiar when changing the incident angle of the shock\cite{vuillon1995reconsideration,chpoun1995numerical,ivanov2001flow}, and is explained with minimal viscous dissipation (MVD) theorem by Hu et al.\cite{hu2021mechanism}. More recently, a separation/attachment hysteresis induced by variation of angle of attack is discovered via numerical simulation in curved compression ramp(CCR)\cite{hu2020bistable}.  


As a classic configuration of shock wave/boundary layer interaction (SBLI)\cite{babinsky2011shock}, compression ramp flow has been studied intensively. Many experiments, numerical simulations, and theories have been performed to gain insights into the effect of adverse pressure gradient (APG)\cite{spina1994physics,fernando1990supersonic}, bulk flow compression\cite{bradshaw1974effect} and centrifugal instabilities in CCR\cite{saric1994gortler,smits2006turbulent}. 
Direct Numerical Simulations (DNS) conducted by Tong et al.\cite{tong2017direct} and Sun et al.\cite{sun2019turbulence} show that the inhomogeneity of the skin friction coefficient is induced by G\"{o}tler-like vortices\cite{ren2015secondary}, and can be intensified downstream that promotes flow transition. It is important to note that Tong et al.\cite{tong2017direct} have also found that CCR flow can be in separation and attachment at two instantaneous time, revealing the existence of two unstable states in turbulence cases.

The separation and attachment states appearing in CCR flow are related to the APG resistance of the boundary layer and the APG induced by the separation and curved wall closely. Spina et al.\cite{spina1994physics} and Jayaram et al.\cite{jayaram1987response} found that APG induced by the curved ramp stimulates turbulence fluctuations and destabilizes the boundary layer. Donovan et al.\cite{donovan1994structure} and Wang et al.\cite{wang2016experimental} discovered that APG distorts the boundary layer structure severely. The boundary layer's resistance to APG is affected by $Ma_{\infty}$ and $T_{w}$ \cite{kubota1967experimental,holden1972shock,spaid1972incipient} and characterized by incompressible shape factor $H_{in}$. Wenzel et al.\cite{wenzel2018dns,wenzel2019self,gibis2019self} found that $H_{in}$ become larger with stronger APG. The APG induced by the separation has been provided by the free interaction theory (FIT)\cite{chapman1958investigation,erdos1962shock,dolling2001fifty}, triple-deck theory\cite{stewartson1969self,neiland1970asymptotic} and MVD\cite{hu2020prediction}. FIT was established by Chapman et al.\cite{chapman1958investigation,erdos1962shock} argues that the supersonic flow is influenced by the local boundary layer and the inviscid contiguous stream rather than the further development of the interaction, supplying a uniform streamwise variation of normalized pressure. By minimizing the shock dissipation among all possible configurations, MVD\cite{hu2020prediction} provides accurate predictions of the shock configurations and the corresponding pressure plateaus and peaks.

Herein, separation hystereses of CCR flow induced by $Ma_{\infty}$-variation and $T_{w}$-variation are studied numerically with 2D simulations. The numerical method is introduced in the second section. The hysteresis processes and corresponding wall quantities are analyzed in the third section. In the fourth section, we discuss the mechanism of the formation of hysteresis theoretically and validate with numerical data. The distributions of sonic line and $H_{in}$ are provided to support the analysis.

\section{Numerical Simulation \label{sec:Num}} 
\subsection{Govern equation and numerical methods} 
We apply DNS to study the bistable states and separation hysteresis in CCR. The non-dimensionalized conservative forms of the continuity, momentum and energy equations in curvilinear coordinates are considered,
\begin{equation} \label{eq:NS1}
\frac{\partial U}{\partial t}+\frac{\partial F}{\partial \xi}+\frac{\partial G}{\partial \eta}=0
\end{equation}
where $ U =J\{\rho, \rho u, \rho v, \rho e\}$ denotes the conservative vector flux, with $\rho$ the density, $e$ the energy per volume, $(u,v)$ velocity components of streamwise and vertical directions, respectively. $J$ is the Jacobian matrix transforming Cartesian coordinates $(x,y)$ into computational coordinates $(\xi, \eta)$. $F$ is the flux term in $\xi$ direction as

\begin{equation} \label{eq:NS-F }
F =F_{c} + F_{v}=J r_{\xi} \left[\begin{array}{c}
\rho u^{*} \\
\rho uu^{*}+ p s_{x}\\
\rho vu^{*} + p s_{y} \\
(\rho e+p) u^{*}
\end{array}\right]  
- J r_{\xi} \left[\begin{array}{c}
0 \\
s_{x} \sigma_{xx}+s_{y} \sigma_{xy} \\
s_{x} \sigma_{yx}+s_{y} \sigma_{yy} \\
s_{x} \tau_{x}+s_{y} \tau_{y}
\end{array}\right]
\end{equation}
where
\begin{equation}
\begin{aligned}
&u^{*}=u s_{x}+v s_{y},\quad s_{x}=\xi_{x} / r_{\xi},\quad \tau_{x}=u \sigma_{xx}+v \sigma_{yx}-q_{x},\\
&r_{\xi}=\sqrt{\xi_{x}^2 + \xi_{y}^{2} },\quad s_{y}=\xi_{y} / r_{\xi},\quad \tau_{y}=u \sigma_{xy}+v \sigma_{yy}-q_{y},\\
&\sigma_{i j}=2 \mu \left[ {\frac{1}{2}\left( {\frac{{\partial {u_i}}}{{\partial {x_j}}} + \frac{{\partial {u_j}}}{{\partial {x_i}}}} \right) - \frac{1}{3}\frac{{\partial {u_k}}}{{\partial {x_k}}}{\delta _{ij}}} \right],\\
& {q_j} =  - \frac{\mu }{{Pr (\gamma  - 1) M_\infty ^2}}\frac{{\partial T}}{{\partial {x_j}}}, \quad (i,j,k = x,y) \\
\end{aligned}
\end{equation}

In above equations, $F_{c}$ and $F_{v}$ are the convective and the viscous terms, respectively. The flux term $G$ in $\eta$ direction has similar forms as $F$. Pressure $p$ and temperature $T$ fulfill $p = \rho T / \gamma Ma_{\infty}^{2}$ with the ratio of specific heats $\gamma = 1.4$ and inflow Mach number $ Ma_{\infty}$, where subscript `$\infty $' represents the free-stream conditions. $\rho, u, v, p, T$ are normalized by the free-stream quantities. The Reynolds number per imeter $Re_{\infty}$ ($m^{-1}$) = $3\times 10^6$ and Prandtl number $Pr = 0.7$. The viscosity $\mu$ fulfills the Sutherland law:$\mu=\frac{1}{Re_{\infty}} \frac{T^{3 / 2}\left(1+T_{s} / T_{\infty}\right)}{T+T_{s} / T_{\infty}}$, where $T_{s} = 110.4K$ and the free-stream temperature $T_{\infty} = 108.1K$.

The present simulations are completing by the in-house code OPENCFD-SC\cite{li2008dns,li2010direct}. The convective flux terms are dealed with Steger-Warming splitting and solved by using WENO-SYMBO method\cite{wu2007direct,martin2006bandwidth}, with a nine-point central stencil and a fourth order accuracy. The viscous flux terms are calculated with eighth-order central difference scheme. Three order TVD-type Runge-kutta method is used for time-advance\cite{gottlieb1998total}. More details about the numerical method can be found in Ref.\cite{li2008dns,li2010direct,martin2006bandwidth}. The code has been validated and successfully applied for many cases, especially for hypersonic compression ramp flows\cite{li2008dns,li2010direct,tong2017numerical,hu2017beta,hu2020bistable,hu2020prediction}. 

\begin{figure}[htbp] 
	\subfigure[\label{subfig:computational_config}]{\includegraphics[width = 0.8 \columnwidth]{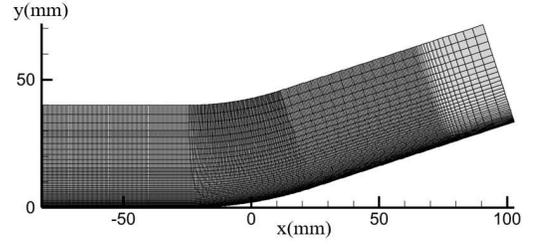}}
	\subfigure[\label{subfig:grid_check}]{\includegraphics[width = 0.8\columnwidth]{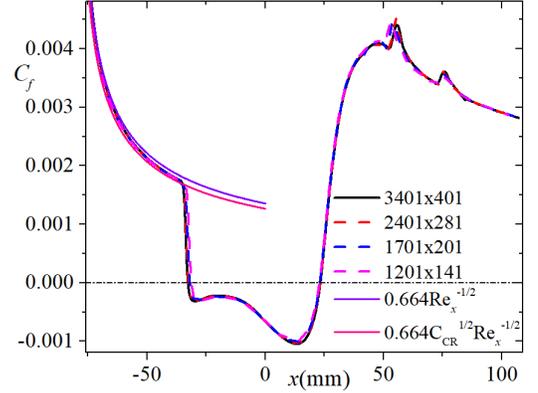}}
	\subfigure[\label{subfig:grid_check2}]{\includegraphics[width = 0.8\columnwidth]{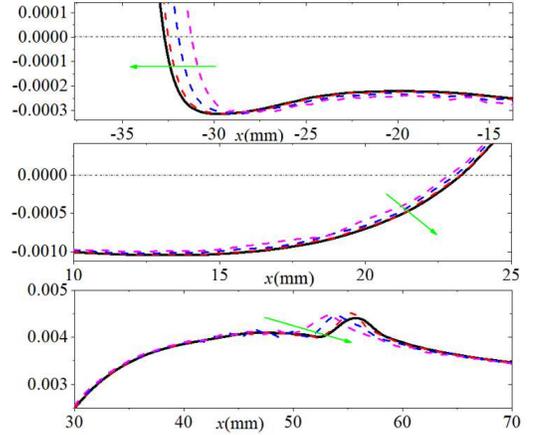}}
	\caption{(a)Sketch of the computational domain and the mesh distribution of grid A (each 10th point is shown). (b)Validation of grid convergence with four grid scales for case S-Ma7.0. The purple and pink solid line are incompressible Blasius solution $C_{f}=0.664Re_{x}^{-1/2}$ and Blasius solution with compressible correction $C_{f}=0.664C_{CR}^{1/2}Re_{x}^{-1/2}$, respectively. (c)Zoom-in plots of the separation, reattachment and pressure peak regions.} \label{fig:numerical}
\end{figure}

\subsection{Computational domain, mesh resolution and flow conditions} 
\begin{table*}[htbp]   
	\caption{Flow conditions for current simulations. }   
	\begin{threeparttable}
		\begin{tabular}{cccc}    
			\toprule    
			$Re_{\infty}$ ($m^{-1}$) \quad & \quad Pr \quad & \quad $Ma_{\infty}$ \quad & \quad $T_{w}$ \\    
			\midrule  
			$3\times 10^{6}$ \quad & \quad 0.7 \quad & \quad 4.5(S), 5.0(A/S), 6.0(A/S), 7.0(A/S), 7.5(A) \quad & \quad 1.5 \\   
			$3\times 10^{6}$ \quad & \quad 0.7 \quad & \quad 6.0 \quad & \quad 1.25(A), 1.5(A/S), 1.75(A/S), 2.0(A/S), 2.25(S) \\  
			\bottomrule   
		\end{tabular}  \label{tab:flow_cond}
		\begin{tablenotes} 
			\item For each case, `A' or `S' represent attachment or separation states, respectively. 
		\end{tablenotes} 
	\end{threeparttable}
\end{table*}

Fig.\ref{subfig:computational_config} shows the computational domain. The streamwise region is $-82mm \leq x \leq 107.2mm$ (the flat plate starts at $x=-80mm$, and $x=0$ is the virtual intersection point of the extension line of the flat plate and the tilt plate). The normal extent of the computational domain is $0mm \le y_{n} \le 40 mm$, $y_{n}$ is the distance to the wall. The turning angle is $\phi = 18^{\circ}$. The curved wall is an arc with the radius of curvature $R=157.84mm$ which tangents with the flat plate and ramp plate (the arc starts at $x=-L$,$L=R\sin\frac{\phi}{2}=25mm$).
 
\par The computational mesh is generated algebraically with $3401 \times 401$ nodes in streamwise and wall-normal direction, respectively, as shown in Fig.\ref{subfig:computational_config}.  The streamwise grid is uniformly spacing with width $0.05mm$ in front of the sponge region. The sponge region contains 160 nodes in streamwise direction with stretch factor 1.015. The wall-normal grid points are clustered in near wall region and the first grid height is $0.01mm$. The laminar boundary layer thickness $\delta$ at $x=-50mm$ is about $0.5mm$, resulting in 41 points inside the boundary layer to guarantee the well-resolutions. The boundary condition of ramp surface is no-slip with constant wall temperature $T_{w}$. The inlet flow is a uniform flow at $x=-82mm$. 

\par To study the bistable states and hysteresis induced by $Ma_{\infty}$-variation and $T_{w}$-variation, two series DNS are conducted. For $Ma_{\infty}$ variation, $Ma_{\infty}$ is set to be 4.5, 5.0, 6.0, 7.0 and 7.5 with fixed $T_{w}=1.5$. For $T_{w}$-variation, $T_{w}$ is set to be 1.25, 1.5, 1.75, 2.0 and 2.25 with fixed $Ma_{\infty}=6.0$. The cases $Ma_{\infty}=5.0,6.0,7.0$ and $T_{w}=1.5, 1.75, 2.0$ are all have two states -- separation states and attachment states. The summary of the flow conditions for current simulations is present in Tab.\ref{tab:flow_cond}.

\subsection{Grid convergence examination} 
\par The mesh convergence with four grid scales, i.e.,$3401 \times 401$(grid A),  $2401 \times 281$(grid B), $1701 \times 201$(grid C) and $1201 \times 141$(grid D), for the separation case $Ma_{\infty}=7.0$ with $T_{w}=1.5$ is studied. Skin friction coefficient distributions $C_{f}$ are shown in Fig.\ref{subfig:grid_check}, where $C_{f}=\frac{2\tau_{w}}{\rho_{\infty} u_{\infty}^2}$, $\tau_{w}=\mu_{w} \frac{\partial u}{\partial y_{n}}|_{y_{n}=0}$ the wall stress. For the attached laminar boundary layer, i.e., $x<38mm$, all $C_{f}$ locate between the incompressible Blasius solution $C_{f}=0.664Re_{x}^{-1/2}$ and Blasius solution with compressible correction $C_{f}=0.664C_{CR}^{1/2}Re_{x}^{-1/2}$, where $C_{CR}=\rho_{w}\mu_{w}/(\rho_{\infty}\mu_{\infty}) \approx T_{w}^{-1/3}$ is Chapman-Rubesin constant\cite{white2006Viscous}. Zoom-in views of separation, reattachment and pressure peak regions are shown in Fig.\ref{subfig:grid_check2}. The two coarsest grids (grid C and D) have some deviations, especially the separation points' location. The separation point of grid D moves 1.5mm downstream, compared with grid A. $C_{f}$ distributions collapse well for the two largest grid scale (grid A and B), indicating the grid size of grid A is applicable.


\section{Result \label{sec:res}} 

\subsection{The hysteresis loops induced by inflow Mach number variation and wall temperature variation}

\begin{figure*}[htbp]
		
	\hspace{2mm}\subfigure[\label{subfig:Ma_loop}]{\includegraphics[width = 2.0\columnwidth]{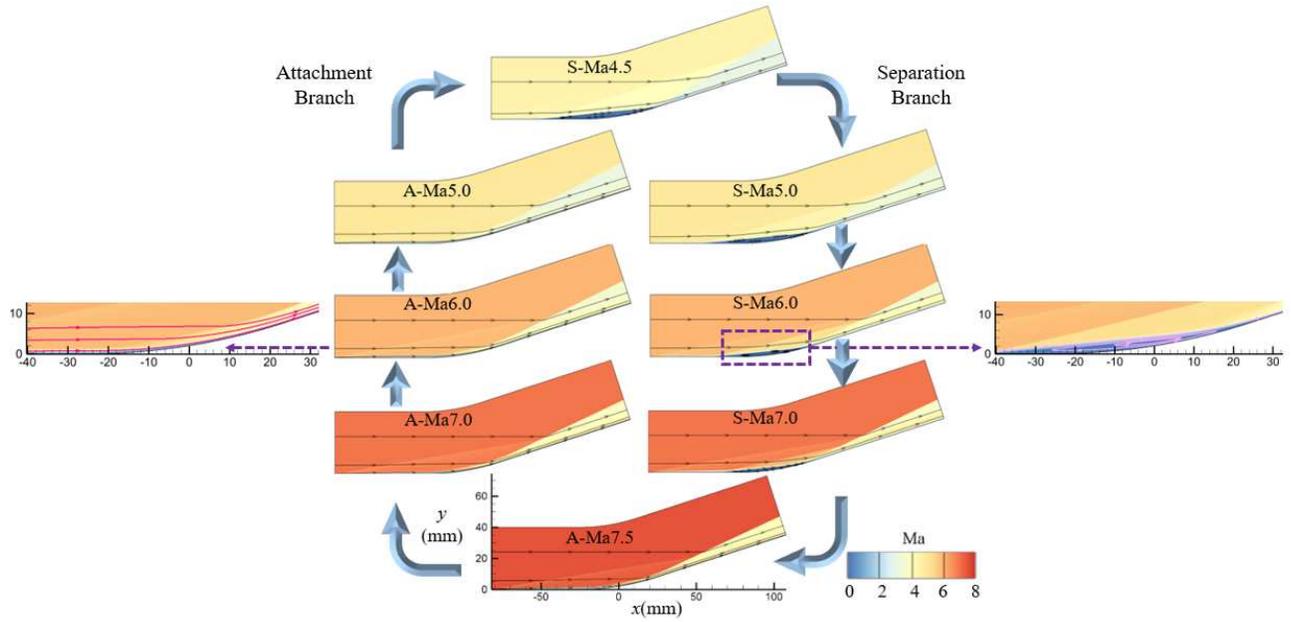}}	
	\hspace{-10mm}\subfigure[\label{subfig:Tw_loop}]{\includegraphics[width = 2.0\columnwidth]{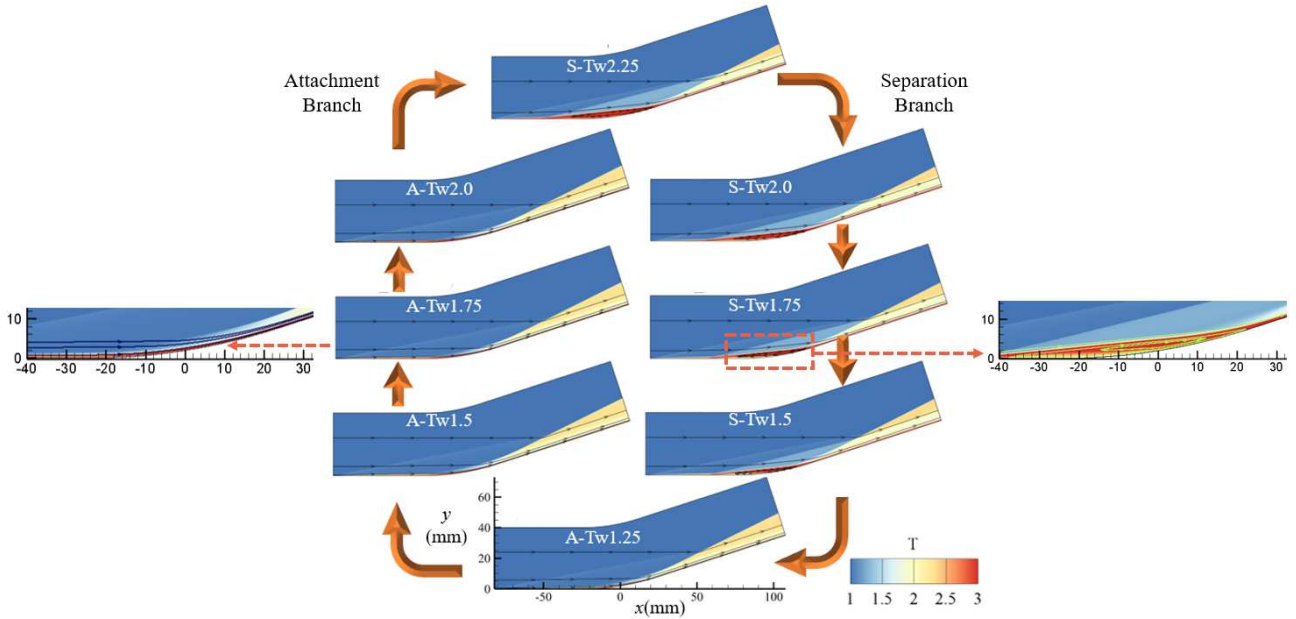}}
	\caption{Separation/attachment hysteresis loops induced by (a) $Ma_{\infty}$-variation, colored by local $Ma$; and (b) $T_{w}$-variation, colored by local $T$. The black lines with arrows are the streamlines. The blue regions in (a) and red regions in (b) are the separation bubbles, as their local $Ma$ are low and $T$ are high compared with inflow. The state `A-Ma5.0' means the attachment state with $Ma_{\infty}=5.0$.}
\end{figure*}

Fig.\ref{subfig:Ma_loop} shows the separation/attachment hysteresis loop induced by the variation of $Ma_{\infty}$ with $T_{w}=1.5$. When $Ma_{\infty}=4.5$, a steady separation flow organized when the supersonic flow passes through CCR. When $Ma_{\infty}$ increases gradually (State S-Ma5.0 to S-Ma7.0 in Fig.\ref{subfig:Ma_loop}), the separation maintains, but the size of separation bubble decreases gradually\cite{green1970interactions}. Here, $Ma_{\infty}$ changes $\Delta Ma_{\infty} = 0.5$ after the flow reaches stable. When $Ma_{\infty}$ increases to 7.5, the separation disappears and the flow reaches attachment state. With $Ma_{\infty}$ starting from 7.5, $Ma_{\infty}$ decreases gradually with the same $\Delta Ma_{\infty}$ after being steady. In a certain range $[Ma_{\infty,s},Ma_{\infty,a}]$ (for the current case, $Ma_{\infty,s} = 5.0$ and $ Ma_{\infty,a} = 7.0$), the attachment state maintains. Once $Ma_{\infty}$ reaches 4.5, the flow will separate rapidly. When $Ma_{\infty} \in [Ma_{\infty,s},Ma_{\infty,a}]$, there will be two different flow states, i.e., separation and attachment, which correspond to the same IBCs, but are determined by different initial conditions.

For the cases of $T_{w}$-variation, the process is similar. Fig.\ref{subfig:Tw_loop} shows the temperature nephogram of steady flow fields corresponding to different $T_{w}$ with $Ma_{\infty}=6.0$. When $T_{w}= 1.25$, the boundary layer attaches to the wall. After the flow becomes steady, the wall is heated with $\Delta T_{w}= 0.25$. It can be found that the flow remains attached till $T_{w}=T_{w,s}=2.0$. When $T_{w}= 2.25$, the flow separates with relatively high temperature fluid filling in the separation bubble\cite{kubota1967experimental,spaid1972incipient}. The separation state maintains but with shrinking separation bubble, even the wall is cooling with $\Delta T_{w}=- 0.25$. Finally, when $T_{w}$ returns to 1.25, the separation bubble will disappear completely.

\begin{figure}[htbp]
	\includegraphics[width = 0.9\columnwidth]{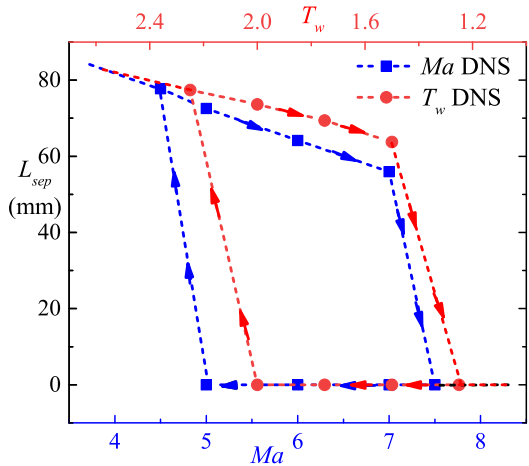}
	\caption{(a)The length of separation $L_{sep}$ in the hysteresis loops induced by $Ma_{\infty}$ and $T_{w}$.} \label{subfig:Lsep_MaTwLoop}
\end{figure}

Fig.\ref{subfig:Lsep_MaTwLoop} shows the hysteresis loops induced by $Ma_{\infty}$ and $T_{w}$ can be characterized by the separation length $L_{sep}= x_{rea}-x_{sep}$, where $x_{rea}$ and $x_{sep}$ are the abscissa of reattachment and separation point, respectively. For different $Ma_{\infty}$ and $T_{w}$, three parameter intervals exist with given certain $Re_{\infty}$ and $R$: 1) overall separation interval (OSI), if the flow can only be in separation; 2) overall attachment interval (OAI), if the flow can only be in attachment; 3) double solution interval (DSI), if both steady attachment and separation states are possible and can exist stably.\cite{hu2020bistable}. By changing the inflow condition ($Ma_{\infty}$) or boundary condition ($T_{w}$), the flow pattern passes through OSI, DSI, OAI, back to DSI sequentially, and finally OSI. This process, therefore, formats the closed separation/attachment hysteresis loop.

\begin{figure}[htbp]
	\includegraphics[width = 0.9\columnwidth]{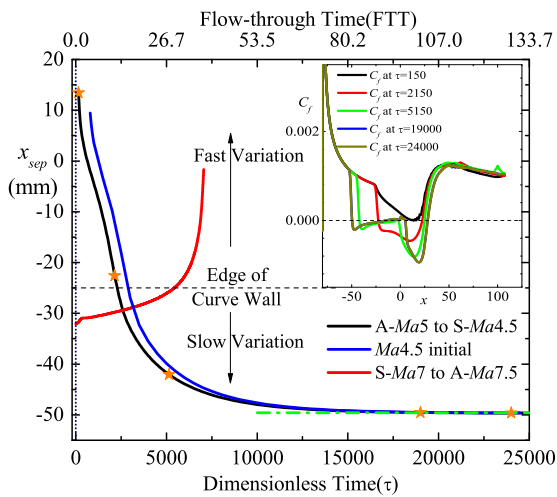}	
	\caption{(b)The separation locations in the transition processes. The black line is the separation locations that the flow transits from $Ma_{\infty} =5.0$ attachment state to $Ma_{\infty}=4.5$. The blue line is separation locations with the flow initialize at $Ma_{\infty}=4.5$. The red line is from $Ma_{\infty} =7.0$ separation state to $Ma_{\infty}=7.5$. The inner figure is the $C_{f}$ distributions at the time labeled with orange stars in the figure.} \label{subfig:Sep_loc}
\end{figure}

\subsection{The attaching and separating processes of boundary layer during the state transition processes} \label{subsec:trans_proc}
We further discuss the transition in OSI $\rightleftharpoons$ DSI. In the marginal cases, the separation and attachment of boundary layer are unsteady. With $Ma_{\infty}$-variation, for example, when $Ma_{\infty}$ decreases from 5.0 to 4.5, the attached boundary layer in attachment state A-Ma5.0 separates to reach the overall separation state S-Ma4.5. The separation process is shown in Fig.\ref{subfig:Sep_loc} as black line. Once the boundary layer on the curved wall receives the disturbance of main flow with lower $Ma_{\infty}$, it crushes and separation starts. In the early stage of separation, the separation point moves upstream rapidly, thus the separation bubble expands dramatically. When the bubble reaches the neighborhood of the edge of the curved wall(i.e., $x=-L=-25mm$), the separation gradually slows down and finally being steady \cite{rudy1991Computation}. It is seen that the transitional process requires about 20000 dimensionless time $\tau$, or equivalently, more than 100 flow-through time (FTT) of the ramp. \footnote{$1\tau$ corresponds to $1/(Ma_{\infty}\sqrt{\gamma R_{g}T_{ref}})\approx \frac{1}{208.4 Ma_{\infty}}$, where $R_{g} = 287 J/(mol \cdot K )$ and $T_{ref}=108.1K$, and the flow passes through 1mm in a dimensionless time. For the separation states, the flow is said to be in convergent when the displacement of separation point is less than 0.01mm in $1000\tau$.} For comparison, we also show the separation process initialized at $Ma_{\infty}=4.5$ (blue line in Fig.\ref{subfig:Sep_loc}). It shows that the flow keeps attaching slightly longer than the process initialized from A-Ma5.0. But once the separation starts, the separation process is consistent with the process starting from A-Ma5.0, and the final separation point converges to the same position. It guarantees the robustness of our simulations, indicating stable states induced by the initial disturbance will not change.

When $Ma_{\infty}$ increases from the separated state of 7.0 to 7.5 (S-Ma7.0 to A-Ma7.5), the attaching process is shown with red line. The initial separation point at $x = - 32.1mm$ gradually moves downstream. When the separation point reaches $x = - 25mm$, the propulsion is accelerated, till the separation bubble completely disappears thus the flow becomes attachment state. Here, the edge of the curved wall seems to have a great influence on the separation and reattachment processes. Once the separation point passes the edge, the reattachment processes will significantly accelerate/decelerate.

During the processes of separating and attaching at the transition states, the relationship among APG induced by the wall, separation bubble and the boundary layer resistance determines which flow pattern will appear.

\subsection{Distributions of skin friction, wall pressure and wall heat flux}

\begin{figure*}[htbp]
	\subfigure[\label{subfig:Cf_AS_Ma}]{\includegraphics[width = 0.66\columnwidth]{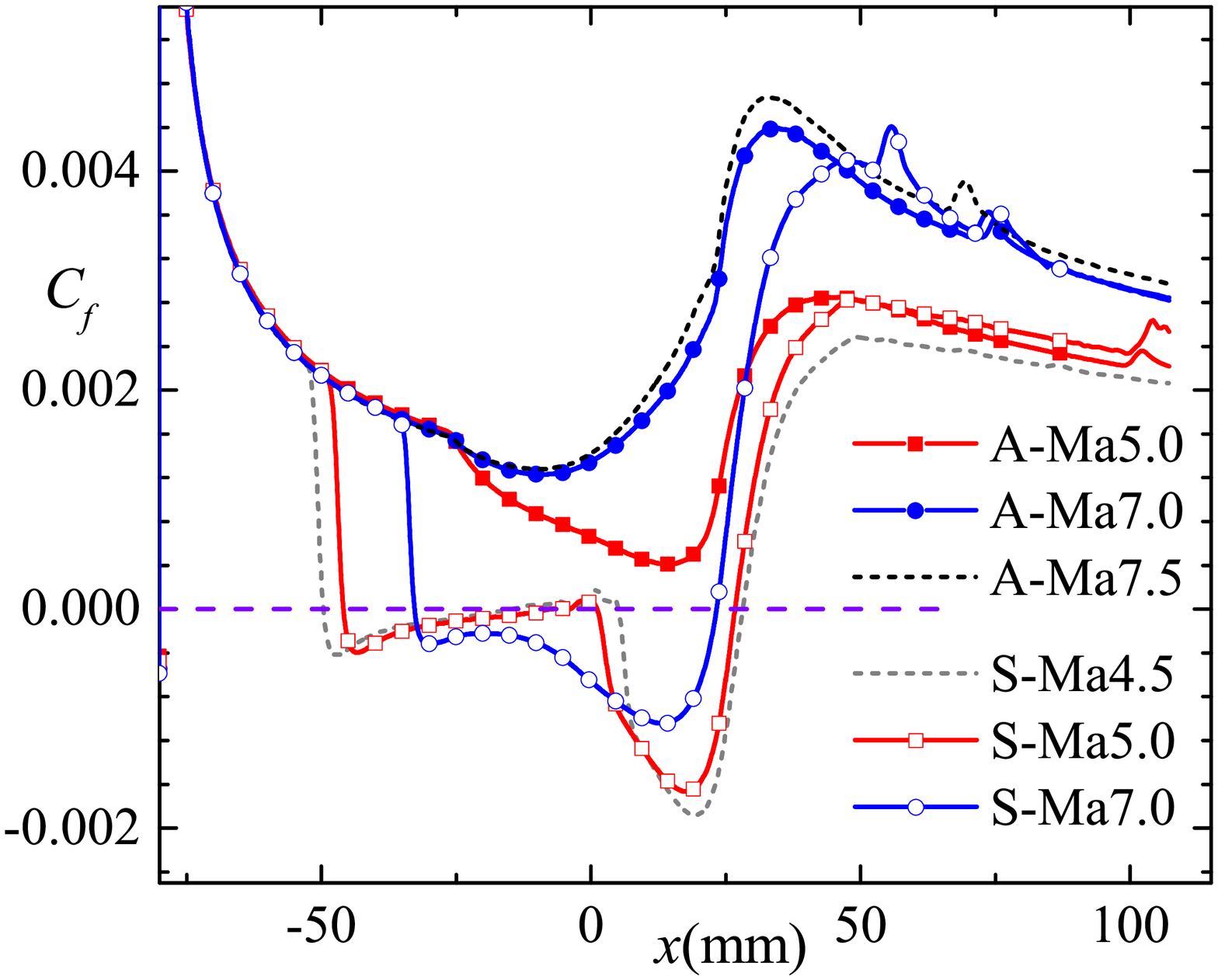}}
	\subfigure[\label{subfig:pw_pinf_AS_Ma}]{\includegraphics[width = 0.66\columnwidth]{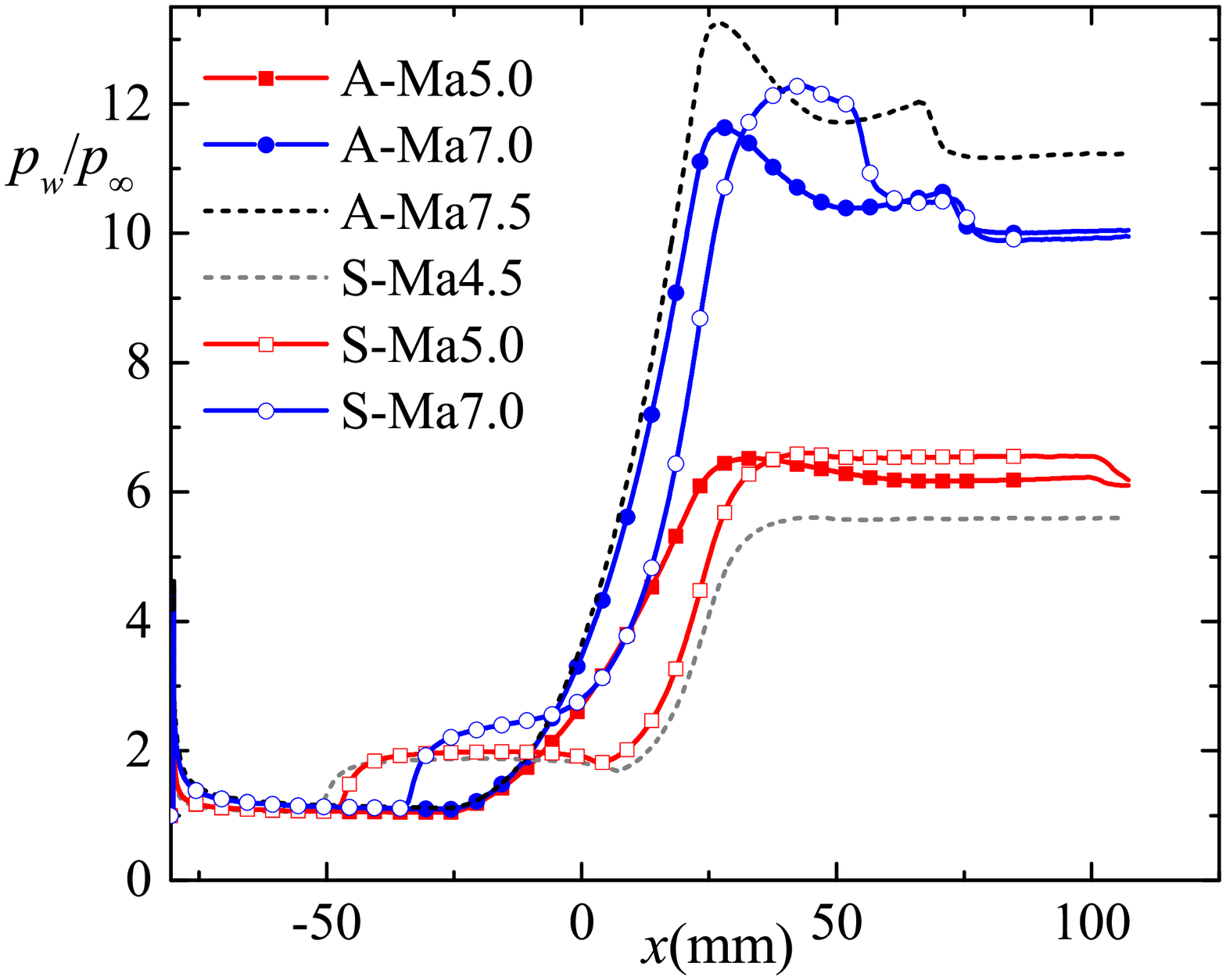}}
	\subfigure[\label{subfig:qw_AS_Ma}]{\includegraphics[width = 0.66\columnwidth]{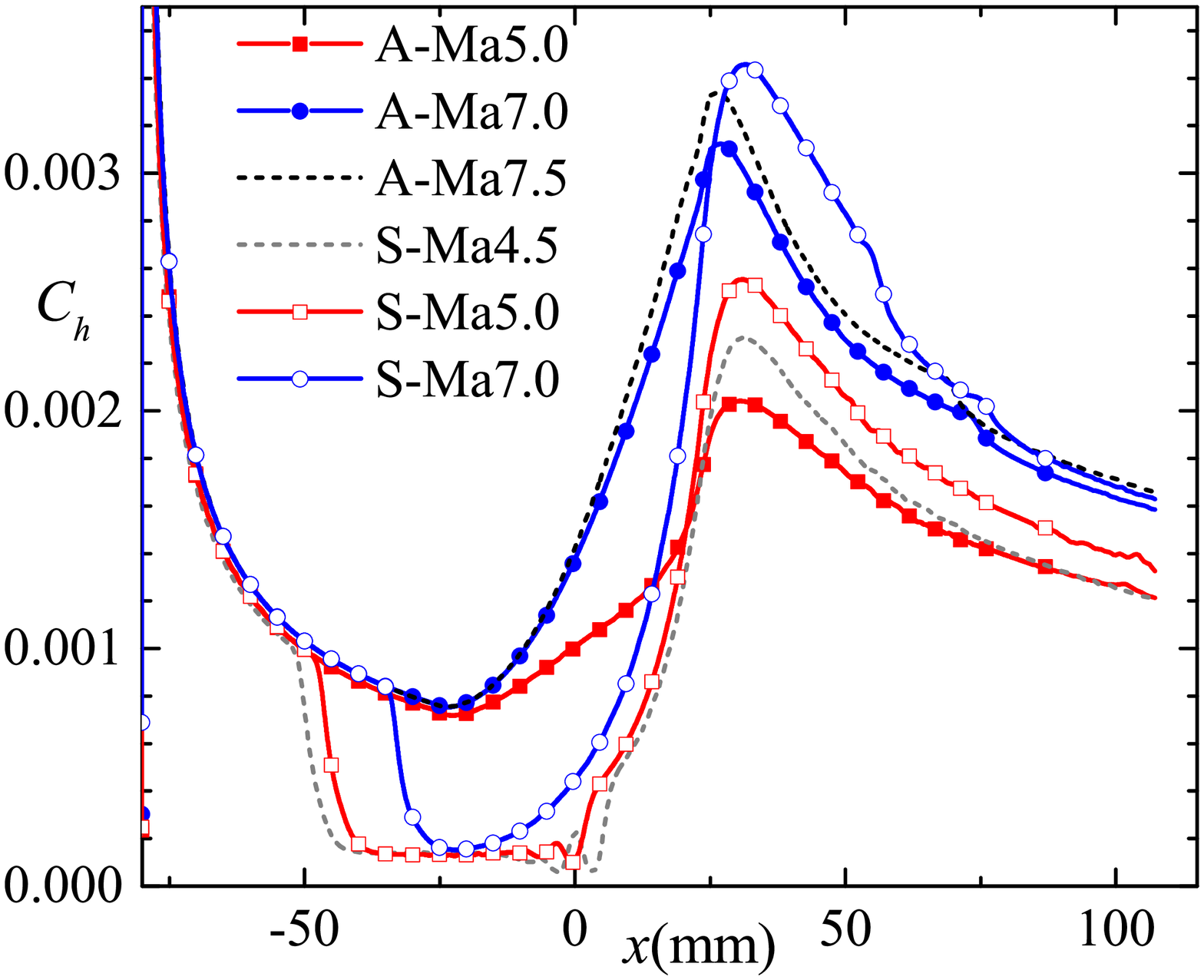}}
	
	\subfigure[\label{subfig:Cf_AS_Tw}]{\includegraphics[width = 0.66\columnwidth]{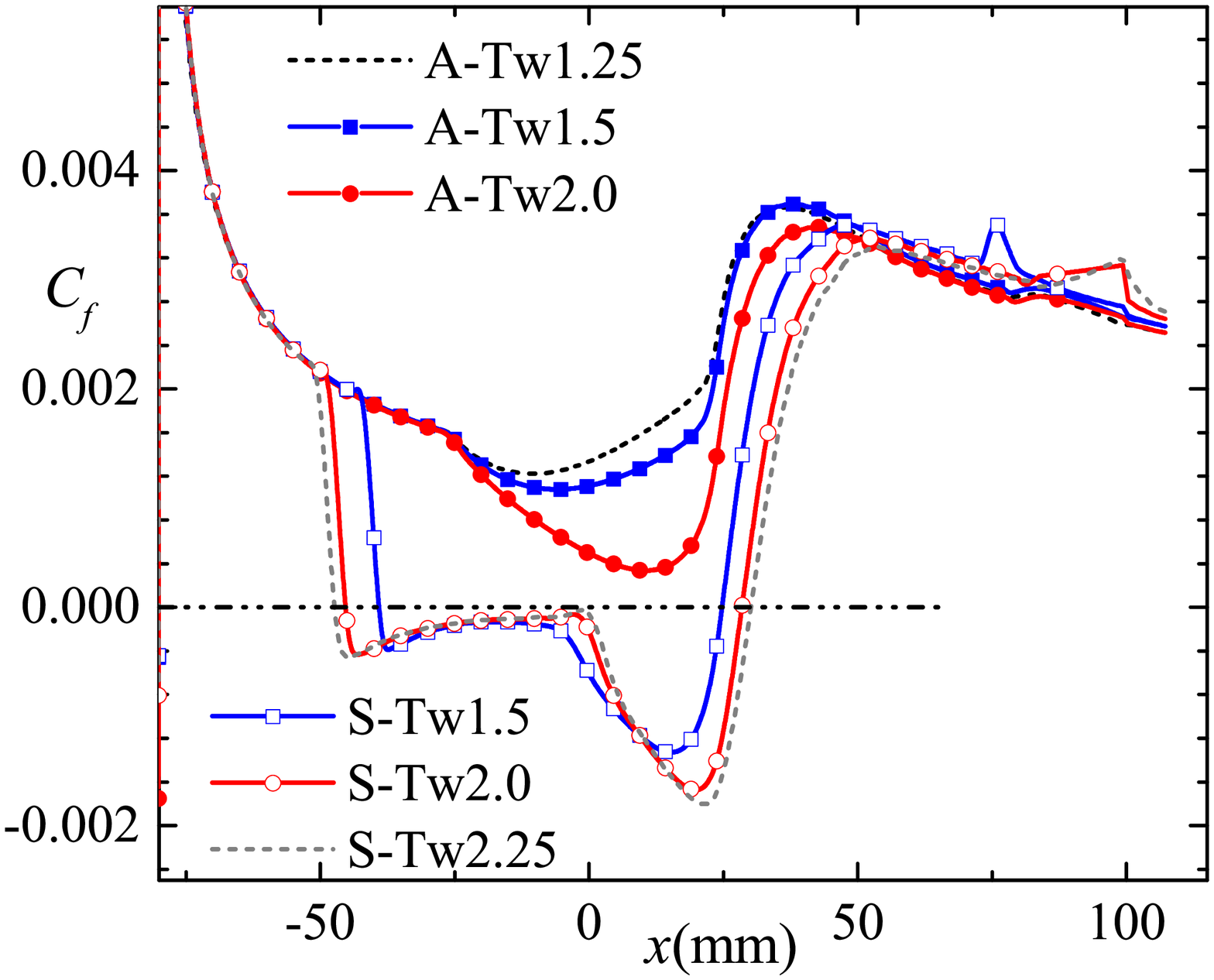}}
	\subfigure[\label{subfig:pw_pinf_AS_Tw}]{\includegraphics[width = 0.66\columnwidth]{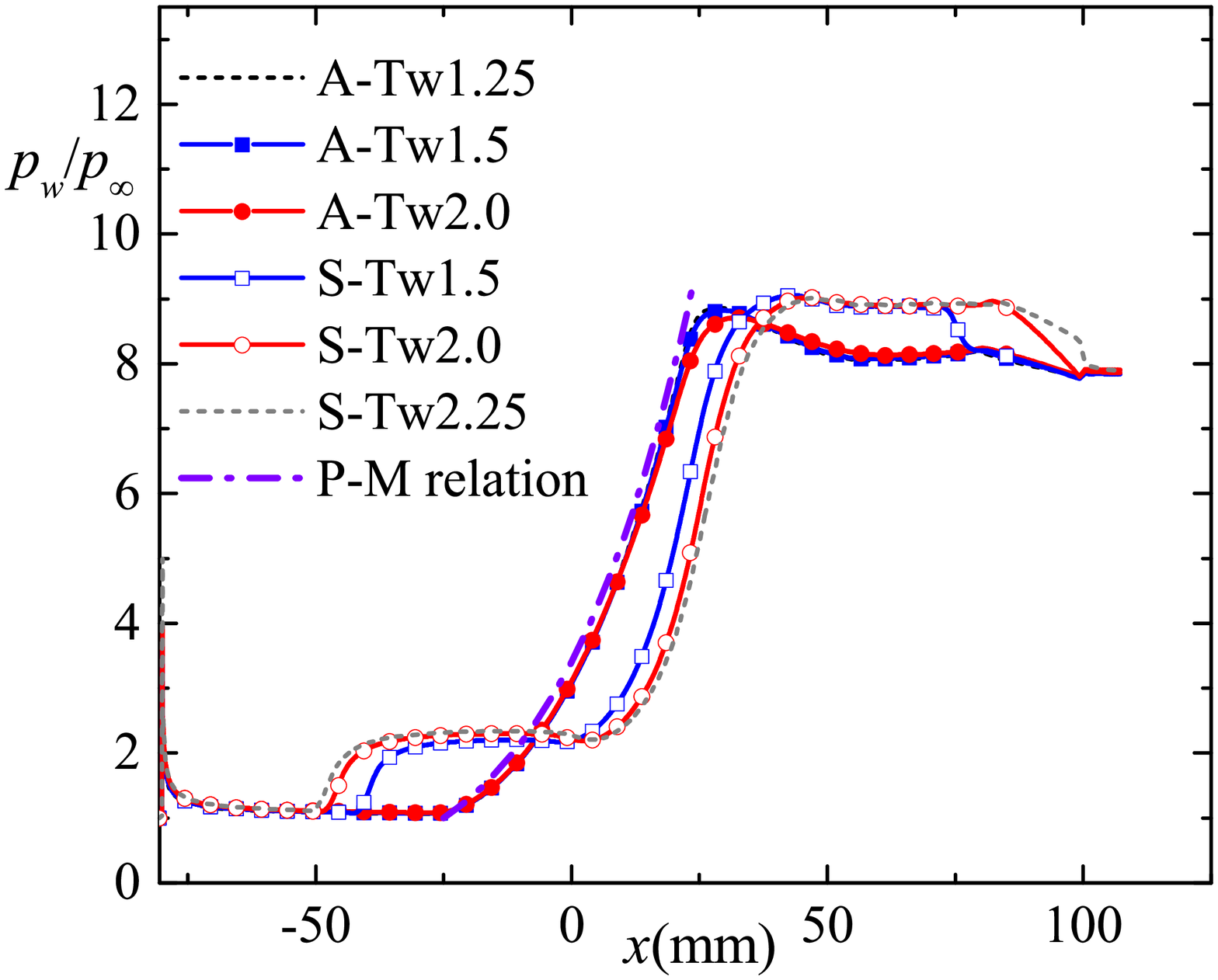}}
	\subfigure[\label{subfig:qw_AS_Tw}]{\includegraphics[width = 0.66\columnwidth]{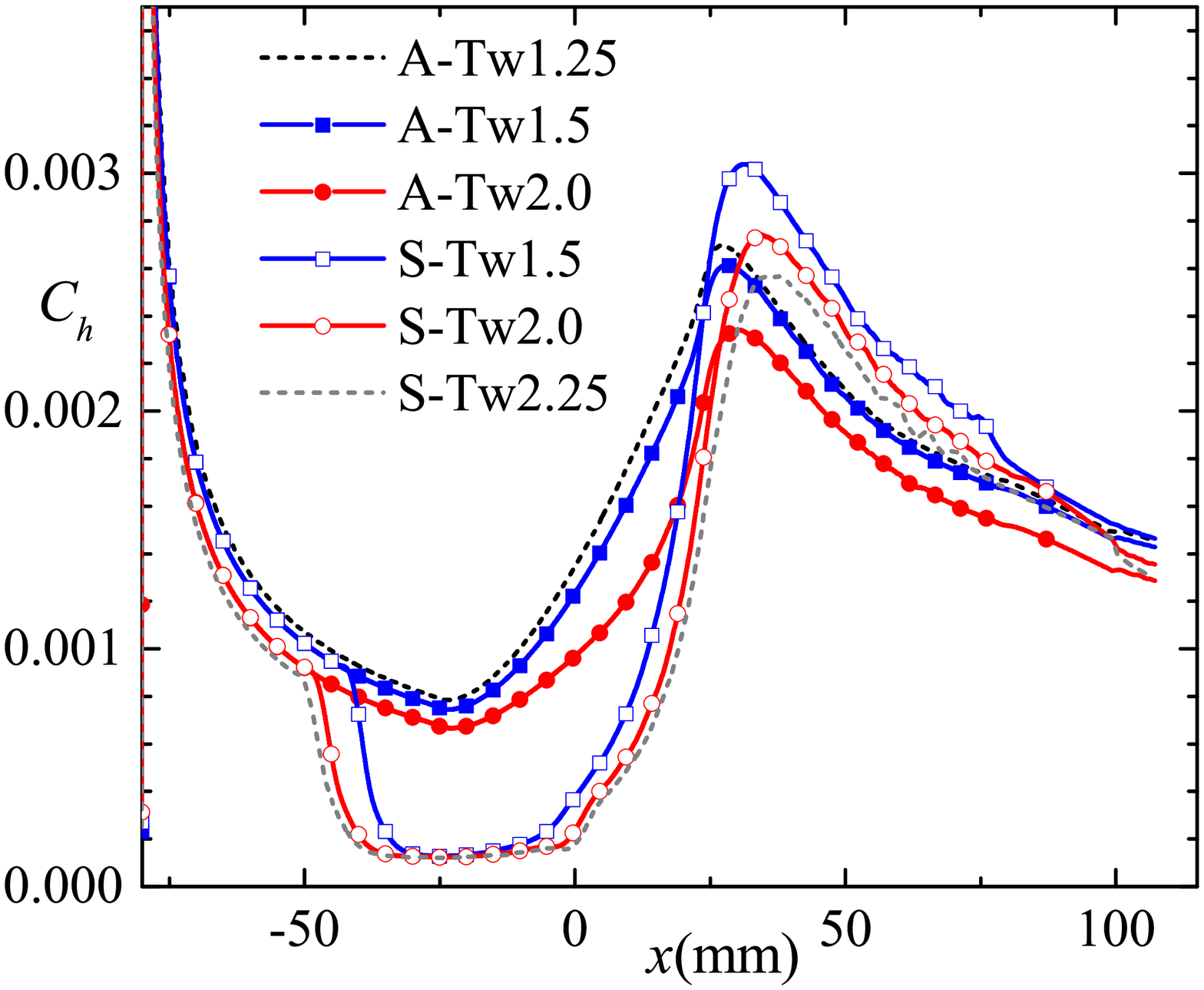}}
	\caption{Skin friction coefficient $C_{f}$, normalized wall pressure $p_{w}/p_{\infty}$, heat flux coefficient $C_{h}$ for attachment and separation states. Upper panel: $Ma_{\infty}$-variation, (a)$C_{f}$; (b)$p_{w}/p_{\infty}$; (c)$C_{h}$. low panel: $T_{w}$-variation, (d)$C_{f}$; (e)$p_{w}/p_{\infty}$; (f)$C_{h}$. The attachment states are represented with solid symbols and the separation states with the same inflow and boundary conditions using empty symbols with the same colors. The black dash lines are states in OAI and the gray dash lines are states in OSI.} \label{fig:CfPwQ}
\end{figure*}

\begin{figure*}[htbp] 
	\subfigure[\label{subfig:field_enlarge_Ma7}]{\includegraphics[width = 0.66\columnwidth]{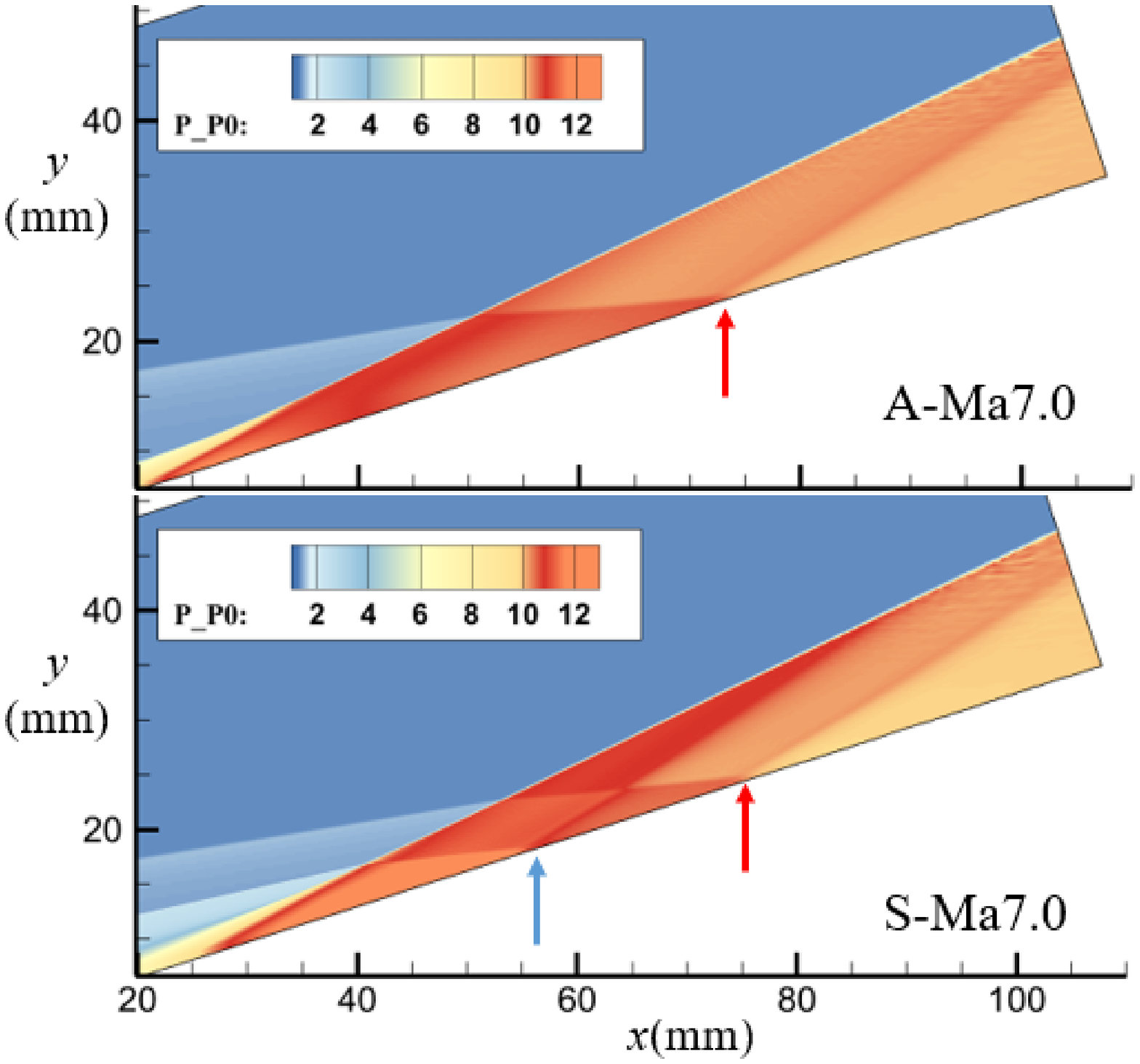}}
	\subfigure[\label{subfig:field_enlarge_Tw2}]{\includegraphics[width = 0.67\columnwidth]{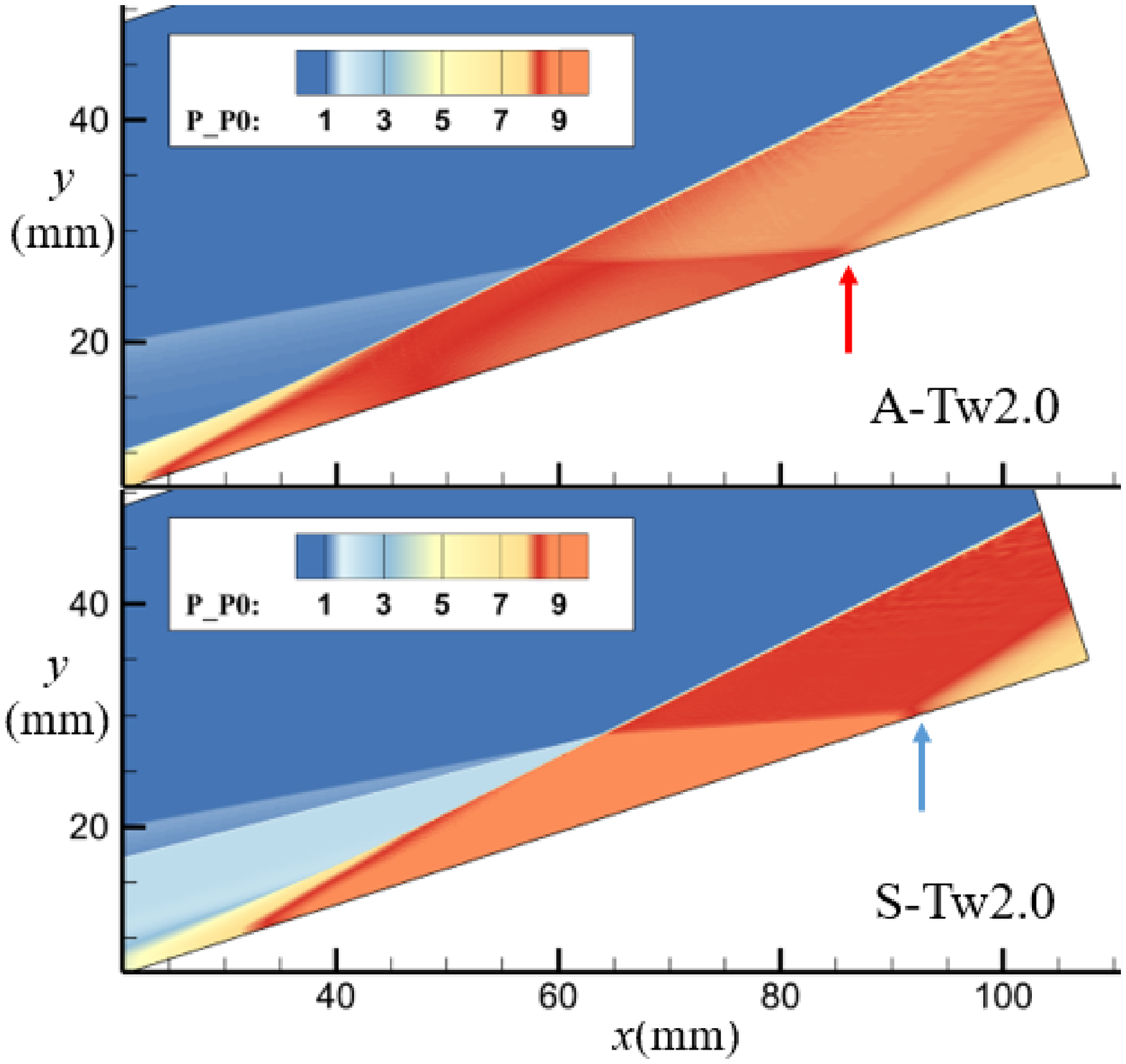}}
	\subfigure[\label{subfig:field_enlarge4}]{\includegraphics[width = 0.48\columnwidth]{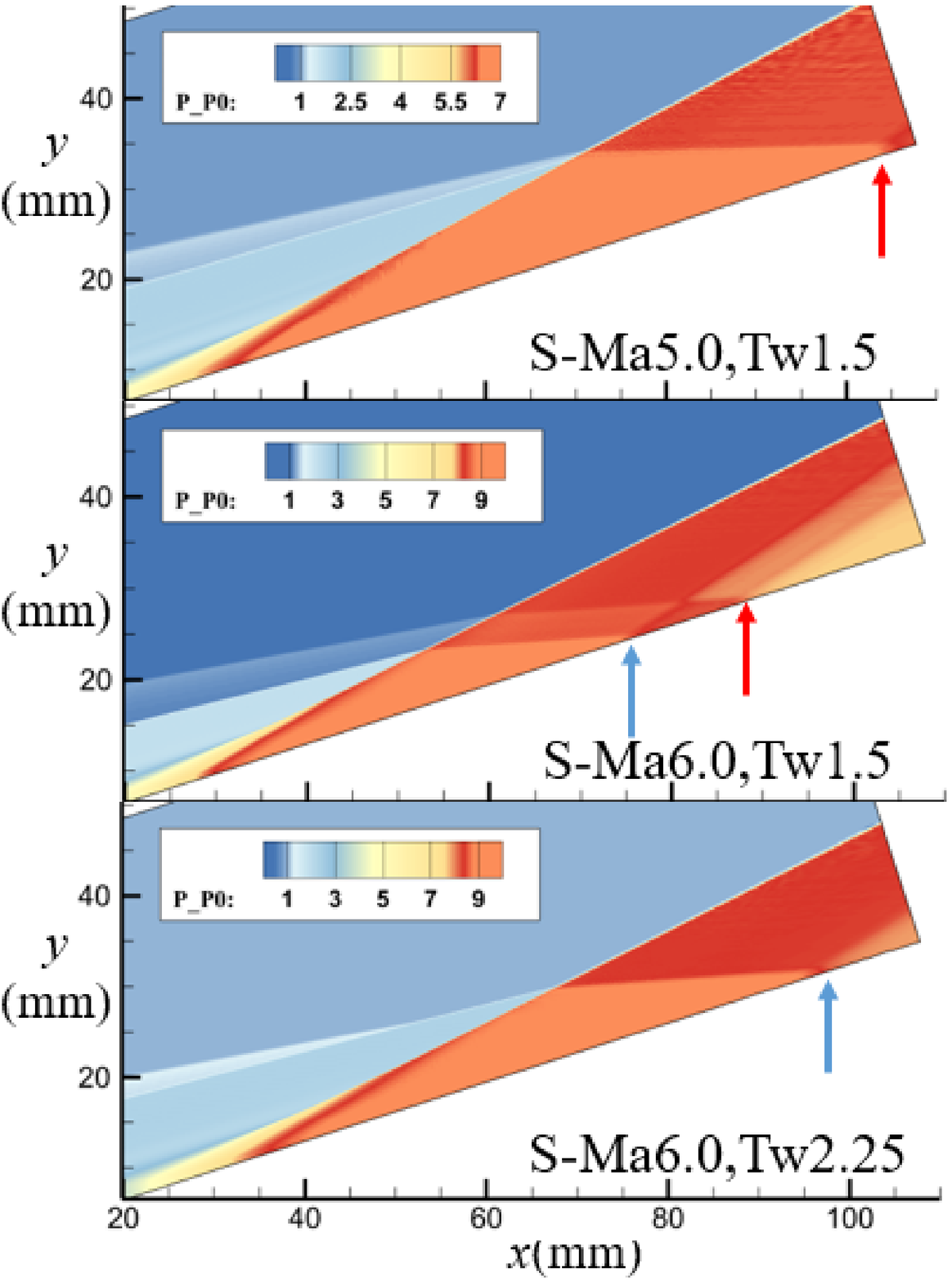}}
	\caption{Zoom-in plots of the pressure nephogram with same inflow and boundary conditions: (a)for $Ma_{\infty}=7.0$, upper panel is attachment state A-Ma7.0, lower panel is separation state S-Ma7.0; (b)for $T_{w}=2.0$, upper panel is attachment state A-Tw2.0, lower panel is separation state S-Tw2.0. (c)upper panel: separation state S-Ma5.0 for $Ma_{\infty}=5.0, T_{w}=1.5$; middle panel: separation state S-Ma6.0 for $Ma_{\infty}=6.0, T_{w}=1.5$; lower panel: separation state S-Tw2.25 for $Ma_{\infty}=6.0, T_{w}=2.25$. The blue and red arrows indicate the origin of pressure drop induced by SSEF and LSEF, respectively.} \label{fig:field_enlarge}
\end{figure*}

The separation and attachment states with the same IBCs correspond to vastly different wall friction $C_{f}$, pressure $p_{w}/p_{\infty}$ and heat flux $C_{h}$ distributions. 

$C_{f}$ distributions with different $Ma_{\infty}$ are shown in Fig.\ref{subfig:Cf_AS_Ma} and different $T_{w}$ in Fig.\ref{subfig:Cf_AS_Tw}. The distributions of attachment states are non-monotonic distributions for $x \in [-25mm,25mm]$ resulting by two competitive effects: the compression effect and APG effect when the flow sweeps the curved wall. The compression effect aggravates the friction while APG effect weakens it. No matter for attachment or separation state, the lower $Ma_{\infty}$ (or higher $T_{w}$) is, the lower $C_{f}$ minima appears, indicating a higher tendency to separate. For a separation state with lower $Ma_{\infty}$ (or higher $T_{w}$), the separation point will move upstream, so as to obtain greater friction to cope with APG, corresponding to a larger separation bubble.

In Fig.\ref{subfig:Cf_AS_Ma}, there are single or double peak(s) of $C_{f}$ downstream of the flow, depending on the state is attachment or separation. They are caused by favor pressure gradient (FPG) induced by the expansion fan interacting with boundary layer, shown in Fig.5. For the mutual peaks of separation states and attachment states, expansion fan is induced by the leading shock \cite{babinsky2011shock} (LSEF, pointed out by red arrows in both panels of Fig.\ref{subfig:field_enlarge_Ma7}.) interacting with the attaching shock. The distinguish peak's expansion fan (separation shock wave induced expansion fan, SSEF, pointed out by blue arrow in lower panel of Fig.\ref{subfig:field_enlarge_Ma7}.) is induced by interaction of separation shock  and attaching shock. With $Ma_{\infty}$ decreasing, the separation bubble becomes larger and the separation shock angle increases, the relative location of leading shock and separation shock will affect the appearance of LSEF, shown in Fig.\ref{subfig:field_enlarge4}. If $Ma_{\infty}$ is small enough, the leading shock wave will interact with the separation shock, thus the LSEF and the corresponding peak in $C_{f}$ will disappear. Fig.\ref{subfig:field_enlarge_Ma7} and Fig.\ref{subfig:field_enlarge_Tw2} show that obvious different processes of shock wave/boundary layer interaction between the separation state and attachment state, even with the same IBCs.

The wall pressure distributions $p_{w}/p_{\infty}$ are shown in Fig.\ref{subfig:pw_pinf_AS_Ma} and Fig.\ref{subfig:pw_pinf_AS_Tw}, where $p_{\infty}=1/(\gamma Ma_{\infty}^2)$. The attached states sustain the isentropic compression process induced by the curved wall\cite{hu2020bistable,zhang2020hypersonic}. The pressure grows with $Ma_{\infty}$ increasing and is independent of the wall temperature, fulfilling
\begin{equation} \label{eq:PM}
	\begin{array}{c}
		\frac{p_{w}(\varphi)}{p_{\infty}} = \left\{ \frac{\vartheta(Ma_{\infty})}{\vartheta \left[ Ma(\varphi) \right]} \right\}^{\frac{\gamma}{\gamma -1}} \\
		\vartheta(Ma) = 1 + \frac{\gamma -1}{2}Ma^2
	\end{array}
\end{equation}
where $Ma(\varphi)$ is obtained with Prandtl-Meyer (P-M) relationship  \cite{anderson2011fundamentals}, and the turning angle at $x$ is
\begin{equation}
\varphi = \arcsin \left( \frac{x+L}{R} \right) \in [0, \phi]  \label{eq:phi}
\end{equation}


In the downstream of the curved wall, wall pressure drops twice both for attachment and separation states downstream of the flow with different mechanism. The first pressure drop of the attachment states is due to the pressure mismatching after isentropic compression overshoot, which is a special phenomenon of CCR flow, compared with the compression flat ramp flow. For the separated states, the SSEF impinging onto the wall is the major reason\cite{babinsky2011shock}, shown in the lower panel of Fig.5(a) and (b). The second drop for attachment states and separation states are both due to the LSEF discussed before. After above two drops, both the separation and attachment states' $p_{w}$ reach the inviscid pressure rise.

The wall heat flux coefficient $C_{h}$ distributions are shown in Fig.\ref{subfig:qw_AS_Ma} and Fig.\ref{subfig:qw_AS_Tw}, where $C_{h}=\frac{2q_{w}}{\rho_{\infty} U_{\infty}^{3}}$, $q_{w}$ is the wall heat flux\cite{gai2019hypersonic}. $C_{h}$ decreases with the development of the inflow boundary layer. For the attachment states, the heat fluxes increase on the curved wall and reach the peak values at the end of it. For the separated states, the slow motion in the separation bubble leading to low heat flux inside the bubble (the fluctuations near $x=0$ are due to the breakdown of vortices, see \cite{gai2019hypersonic,smith1991interactive}), and the peak values occur after the boundary layer reattaching. Comparing with the separation states, the peak values of the attachment states with the same IBCs are reduced greatly (up to 26\% reduction in present cases) and the locations are moving upstream.


The distributions of $C_{f}$, $p_{w}/p_{\infty}$ and $C_{h}$ show that although both $Ma_{\infty}$ and $T_{w}$ variations can lead to bistable states and hysteresis loop, the processes are not quiet the same. The distributions of $p_{w}/p_{\infty}$ in separation and post-reattachment regions collapse while $T_{w}$ varies, but not in cases for $Ma_{\infty}$ variation. For the separation states, $Ma_{\infty}$ variation not only changes the size of separation bubble but also the separation angle. However, it seems that the $T_{w}$ variation mainly changes the size of separation bubble but hardly the separation angle (shown in Fig.\ref{subfig:field_enlarge4}), which consists with the theory proposed by MVD.\cite{hu2020prediction}. Thus the pressure plateaus and peaks collapse.

\section{Discussion } \label{sec:dis}
In Ref.\cite{hu2020bistable}, we defined APG raised by separation as $I_{sb}$; the boundary layer's APG resistance as $I_{b}$; APG induced by the curved wall as $I_{cw}$. The authors proposed a criterion to determine the DSI: being in OAI, if $I_{b} > I_{sb}$; being in OSI, if $I_{b} < I_{cw}$; being in DSI, if $I_{cw} < I_{b}$ and $I_{b} < I_{sb}$. As an first attempt to analyze the mechanism, we implied the hypothesis of $I_{cw}<I_{sb}$ and have not considered the spatial distributions of $I_{b}, I_{cw}$ and $I_{sb}$. In this paper, we further states that, the occurrence of double solution depends on the separation and attachment conditions, whose self-consistency is guaranteed by the spatial distributions of $I_{b}, I_{cw}$ and $I_{sb}$. We first propose the stable conditions of separation and attachment in the language of APG, then analyze their variations with $Ma_{\infty}$ and $T_{w}$ variations combining with FIT and isentropic compression results and provide the occurrence mechanism of double solution and hysteresis. Finally, We validate our theory with numerical data.

\subsection{Stable conditions of separation and attachment in the language of APG}
Prop.\ref{pro:att} and Prop.\ref{pro:sep} are the stable conditions of separation and attachment in the language of APG.
\begin{mypro} [stable condition of attachment]
	 For $\forall x \in L_{x}, I_{b}(x) \geq I_{cw}(x)$, if the boundary layer is stable attachment, where $L_{x}$ is the streamwise domain. \label{pro:att}
\end{mypro}

If Prop.\ref{pro:att} is not held, i.e., there exists a location $x$, s.t. $ I_{b}(x) < I_{cw}(x)$, the APG of curved wall would crush the boundary layer and the stable attachment will no longer hold.

The stable conditions of separation is more complicated. It is based on the fact that, in the separation, the location $x_{max}$ of the maximum APG induced by separation ($I_{sb}(x_{max})$) does not occur at the separation point $x_{sep}$ but in front of it, which implies that $I_{sb}(x_{sep}) \leq I_{sb}(x_{max})$ with $x_{max}<x_{sep}$, where $I_{sb}(x_{sep})$ is the APG at $x_{sep}$. This fact is verified in Fig.\ref{fig:Ib}. Then the stable conditions of separation can be expressed as

\begin{mypro} [stable conditions of separation]
	If the boundary layer is separated at $x_{sep}$, the following two conditions should hold:\\
	(1)For $\forall x \in [x_{max}, x_{sep}], I_{b}(x) \geq I_{sb}(x)$;\\
	(2) $\exists x_{db} \in [x_{max}, x_{sep}], I_{b}(x_{db}) = I_{sb}(x_{db})$.  \label{pro:sep}
\end{mypro}

Here, $x_{db}$ is a dynamical balance point. If the condition (1) of Prop.\ref{pro:sep} is not held, i.e., there exists a location $x_{2}$, s.t., $ I_{b}(x_{2}) < I_{sb}(x_{2})$, the separation APG would push the separation point upstream. If the condition (2) of Prop.\ref{pro:sep} is not held, i.e., for any location $x$, s.t., $ I_{b}(x) > I_{sb}(x)$ (combined with condition (1)), the separation would no longer hold. However, the exact location of $x_{db}$ is not clear in advance, and could vary for different cases. We will see that variability of $x_{db}$ is important for the DSI occurrence mechanism self-consistency.

Both Prop.\ref{pro:att} and Prop.\ref{pro:sep} must be satisfied if DSI occurs in CCR flows whose existence has been demonstrated. The margins of DSI are the parameters that Prop.\ref{pro:att} or Prop.\ref{pro:sep} is no longer satisfied. OAI occurs when Prop.\ref{pro:sep} do not hold, while OSI occurs when Prop.\ref{pro:att} do not hold. The variation of parameters leads to the spatial change of $I_{sb},I_{cw}$ and $I_{b}$, resulting in Prop.\ref{pro:att} and Prop.\ref{pro:sep} satisfied and dissatisfied sequentially, which is the mechanism of the separation hysteresis. Thus how the relative magnitude of $I_{sb},I_{cw}$ and $I_{b}$ varies with $Ma_{\infty}$ and $T_{w}$ variation need concerning.


\begin{figure}[htbp]
	\includegraphics[width = 0.99\columnwidth]{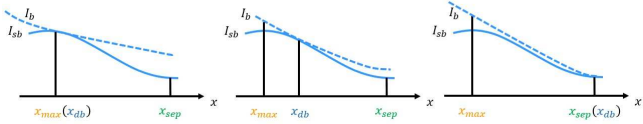} 
	\caption{Possible locations of dynamics balance point $x_{bd}$.} \label{fig:dyna_balance}
\end{figure}

\subsection{The mechanism of hysteresis -- theory analysis} \label{subsec:MechHyster_Theory}
The hysteresis loop induced by $Ma_{\infty}$-variation is considered. For the initial state starts from OSI, i.e., $Ma_{\infty} < Ma_{\infty,s}$, as the initial inflow condition, the boundary layer's APG resistance is too weak (which is characterized by the wall shear, height of local sonic line $y_{c}$ and incompressible shape factor $H_{in}$, discussed in more detail in the following subsections.) to withstand the APG of the curved wall, and the flow separates. Once the reverse flow emerges, the separation bubble will grow gradually and form a virtual wedge, which changes the shock configurations and pressure distributions of the flow field, especially the APG near the separation point\cite{hu2020bistable}.

We first analyze $I_{sb}$ of OSI and DSI. $I_{sb}$ can be estimated as


\begin{equation}
	I_{sb} = \left. \frac{\partial p}{\partial x}\right|_{max,s} \sim \frac{\Delta p}{L_{i}}
\end{equation}
where subscript `$max,s$' represents the maximum in $[x_{oi}, x_{sep}]$, subscript `$oi$' and `$sep$’ represent quantities at origin interaction and separation locations, respectively\cite{chapman1958investigation,erdos1962shock,delery1985shock}. $\Delta p$ is the pressure rise of the separation and $L_{i}$ is the interaction length scale. FIT\cite{chapman1958investigation,erdos1962shock} provides 
\begin{equation}
	\begin{array}{c}
		\Delta p = \frac{0.8}{2} p_{oi}\gamma Ma_{\infty}^2\left( C_{f,oi}^{1/2} \left( Ma_{\infty}^2 -1 \right)^{-1/4} \right) \\
		L_{i} = \delta^{*}_{oi} C_{f,oi}^{-1/2} \left( Ma_{\infty}^2 -1 \right)^{-1/4}
	\end{array}
\end{equation}
where $\delta^{*}$ is the displacement thickness and $p_{oi} \approx 1/(\gamma Ma_{\infty}^2)$. Thus we can obtain

\begin{equation} \label{eq:Isb_FIT}
	I_{sb} \sim \frac{C_{f,oi}}{\delta^{*}}
\end{equation}

\begin{figure}[htbp]
	\centering
	\subfigure[\label{subfig:thick_Ma}]{\includegraphics[width = 0.8\columnwidth]{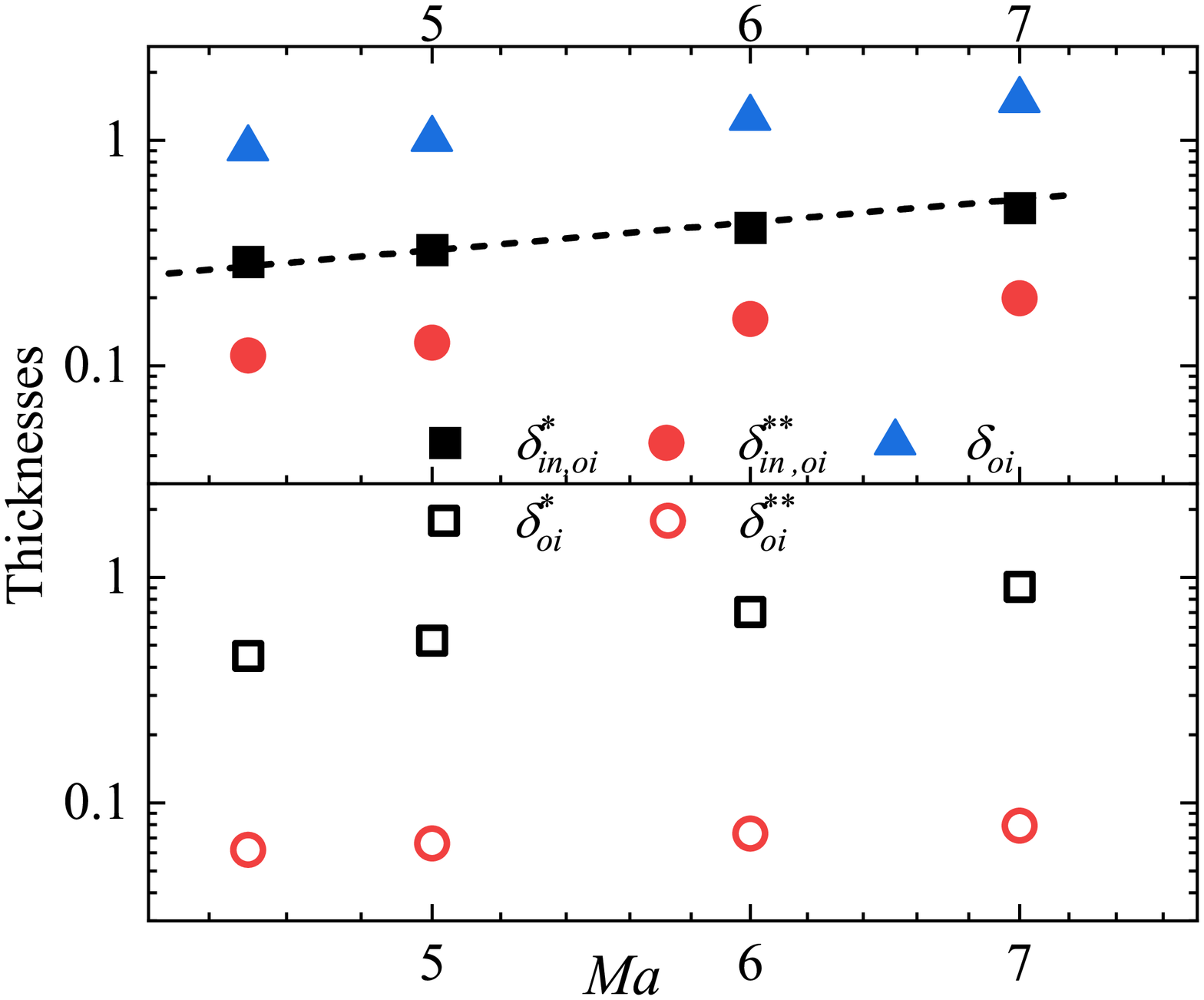}} 
	\hspace{4mm}
	\subfigure[\label{subfig:thick_Tw}]{\includegraphics[width = 0.8\columnwidth]{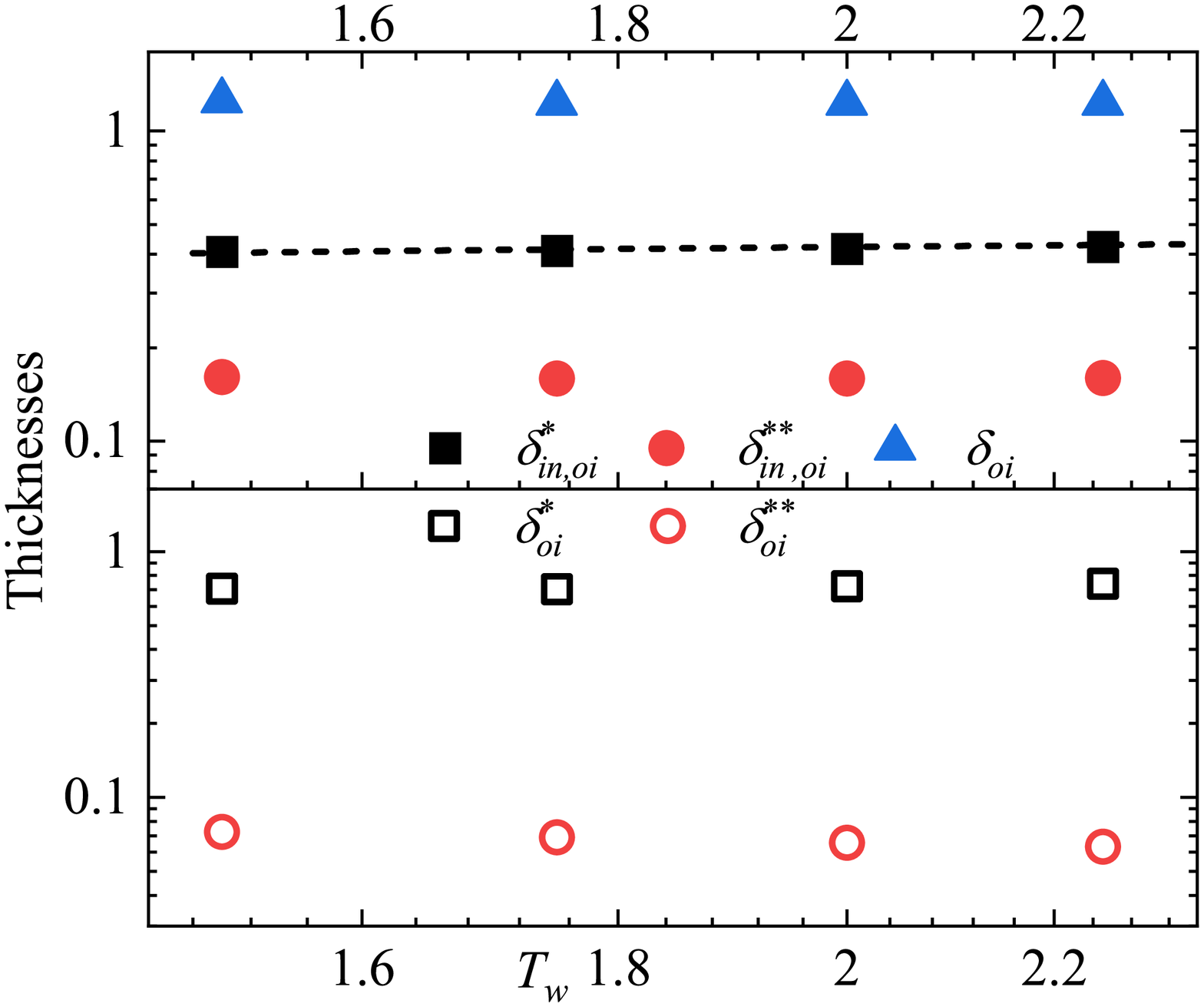}} \hspace{-2mm}
	\caption{Origin interaction locations' boundary layer thicknesses $\delta$, incompressible displacement thicknesses $\delta_{in,oi}^{*}$, incompressible momentum thicknesses $\delta_{in,oi}^{**}$, compressible displacement thicknesses $\delta_{oi}^{*}$ and compressible momentum thicknesses $\delta_{oi}^{**}$ defined as Eq.(\ref{eq:thick}) and Eq.(\ref{eq:com_thick}) \cite{wenzel2018dns} in appendix. (a) $Ma_{\infty}$-variation; (b) $T_{w}$-variation.} \label{fig:Thickness}
\end{figure}

The incompressible displacement thickness at the location of origin interaction $\delta_{in,oi}^{*} = \delta_{in}^{*}(x_{oi})$ can be used to replace $\delta^{*}_{oi}$ by characterizing the kinetic process\cite{wenzel2019self}, where $\delta_{in}^{*}(x)$ is defined in Eq.(\ref{eq:thick}) in Appendix. With $Ma_{\infty}$ increasing, $\delta_{in,oi}^{*}$ increases gradually and empirically fulfills $\delta_{in,oi}^{*} = 0.03 \left( Ma_{\infty}^2 -1 \right)^{0.75 \pm 0.02}$ (shown in Fig.\ref{subfig:thick_Ma}) and $C_{f,oi}$ decreases as $x_{oi}$ moves downstream, which induces the reduction of $I_{sb}$. 

Secondly, We analyze $I_{b}$ of OSI and DSI. $H_{in}$ of the inflow boundary layer decreases (shown in Fig.\ref{subfig:Hfactor_Ma}) with increasing $Ma_{\infty}$, which indicates $I_{b}$ becomes larger before $x_{max}$. When $Ma_{\infty}$ transits from DSI to OAI, i.e., $Ma_{\infty} > Ma_{\infty,a}$, it's expected that $I_{b} > I_{sb}$, the boundary layer resistance is strong enough to withstand the separation APG, thus the separation will disappear. It should be noted that if the spatial message is lost, i.e., $[x_{max}, x_{sep}]$ condenses to one point $x_{0}$, Prop.\ref{pro:sep} indicates $I_{b}(x_{0}) = I_{sb}(x_{0})$ will conflict with the reduction of $I_{sb}$ and increasing of $I_{b}$ with increasing $Ma_{\infty}$.

\begin{figure}[htbp]
	\includegraphics[width = 0.99\columnwidth]{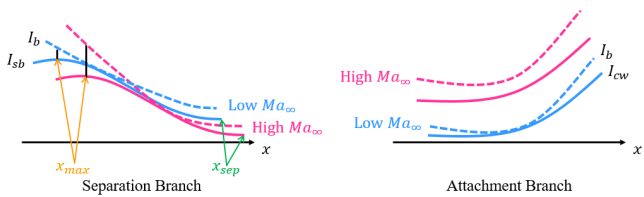} 
	\caption{The schematic magnitude of $I_{sb},I_{cw}$ and $I_{b}$ of $Ma_{\infty}$-variation. Left: separation branch; right: attachment branch.} \label{fig:Ma_var_3I}
\end{figure}

Thirdly, we analyze $I_{cw}$ and $I_{b}$ of OAI and DSI. When $Ma_{\infty}$ starts from OAI, the boundary layer can resist the APG caused by the curved wall. With the definition, it indicates that $ I_{b} \geq I_{cw}$, where $I_{b}$ is the boundary layer's APG resistance. Because the pressure distribution on the curved wall approximately satisfies the isentropic compression process, the wall pressure gradient can be obtained by differentiating the P-M relation

\begin{equation} \label{eq:dPM} 
		I_{cw} = \left. \frac{dp}{dx_{s}} \right|_{max, cw} = p_{\infty} \frac{d}{dx_{s}}\left\{ \frac{\vartheta(Ma_{\infty})}{\vartheta \left[ Ma(x_{s}) \right]} \right\}^{\frac{\gamma}{\gamma -1}}
\end{equation}
where $x_{s}=R\arcsin\left(\frac{x+L}{R}\right)$ is the streamwise abscissa on the curved wall. According to this relation, for a given position, with $Ma_{\infty}$ decreasing, $I_{cw}$ will decrease. Simultaneously, the minimum value of $C_{f}$ on the curved wall decreases with $H_{in}$ increasing (Fig.\ref{subfig:Hfactor_Ma}), which means $I_{b}$ is decreasing. The reduction ratio of $I_{b}$ is faster than $I_{cw}$ with $Ma_{\infty}$ decreasing. When $Ma_{\infty}$ transits from DSI to OSI, there at least one point fulfills $I_{b} < I_{cw}$, the attachment state will not be maintained. The schematic distributions of $I_{sb},I_{cw}$ and $I_{b}$ are proposed in Fig.\ref{fig:Ma_var_3I}.

A similar analysis can be applied for the hysteresis loop induced by $T_{w}$-variation. When the initial state starts from OSI, i.e., $T_w > T_{w,s}$, the boundary layer separates. With decreasing $T_{w}$, $I_{sb}$ estimated by Eq.(\ref{eq:Isb_FIT}) become larger. $I_{b}$ also become larger characterized by large shear(Fig.\ref{subfig:Cf_AS_Tw}), small $H_{in}$(Fig.\ref{subfig:Hfactor_Tw}) and low height of sonic line (Fig.\ref{subfig:sonicline_Tw}). When $T_{w}< T_{w,a}$, $I_{b} > I_{sb}$ thus the boundary layer become attachment.

\begin{figure}[htbp]
	\includegraphics[width = 0.99\columnwidth]{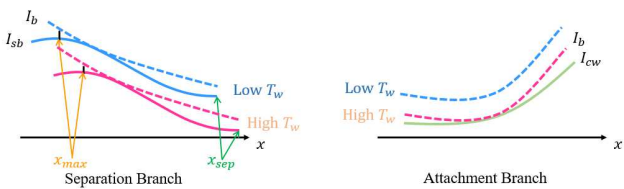} 
	\caption{The schematic magnitude of $I_{sb},I_{cw}$ and $I_{b}$ of $T_{w}$-variation. Left: separation branch; right: attachment branch.} \label{fig:Tw_var_3I}
\end{figure}

When $T_{w}$ starts from OAI, it infers that $I_{b} > I_{cw}$. $I_{cw}$ of isentropic compression is independent with $T_{w}$. With wall heating, the weaker shear(Fig.\ref{subfig:Cf_AS_Tw}), larger $H_{in}$(Fig.\ref{subfig:Hfactor_Tw}) and higher sonic line (Fig.\ref{subfig:sonicline_Tw}) indicate $I_{b}$ becomes smaller. When $T_{w} > T_{w,s}$, $I_{b} > I_{cw}$ no longer holds.

\subsection{The mechanism of hysteresis -- data validation}

\begin{figure}[htbp]
	\subfigure[\label{subfig:Ib_Ma}]{\includegraphics[width = 0.8\columnwidth]{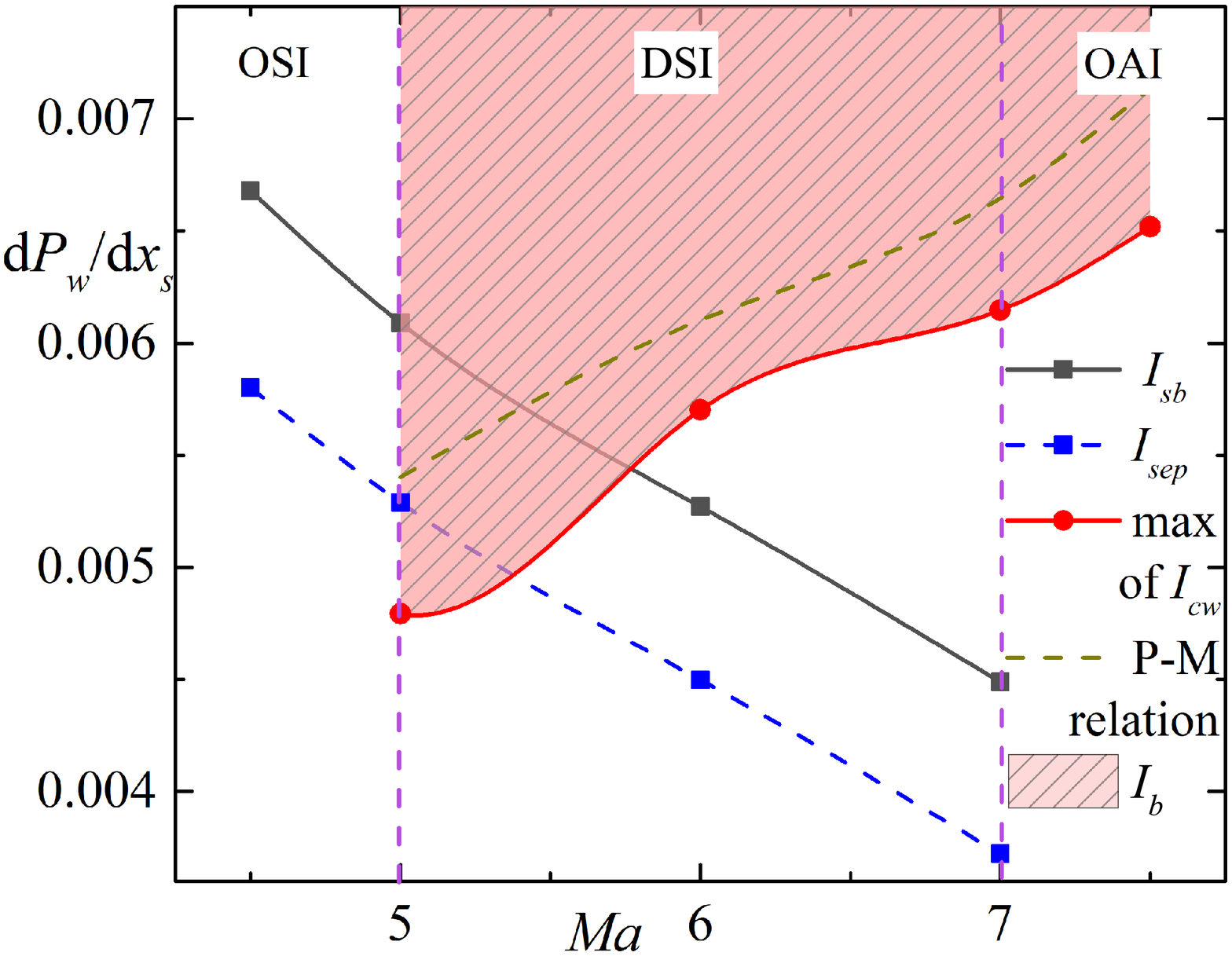}} 
	\hspace{4mm}
	\subfigure[\label{subfig:Ib_Tw}]{\includegraphics[width = 0.8\columnwidth]{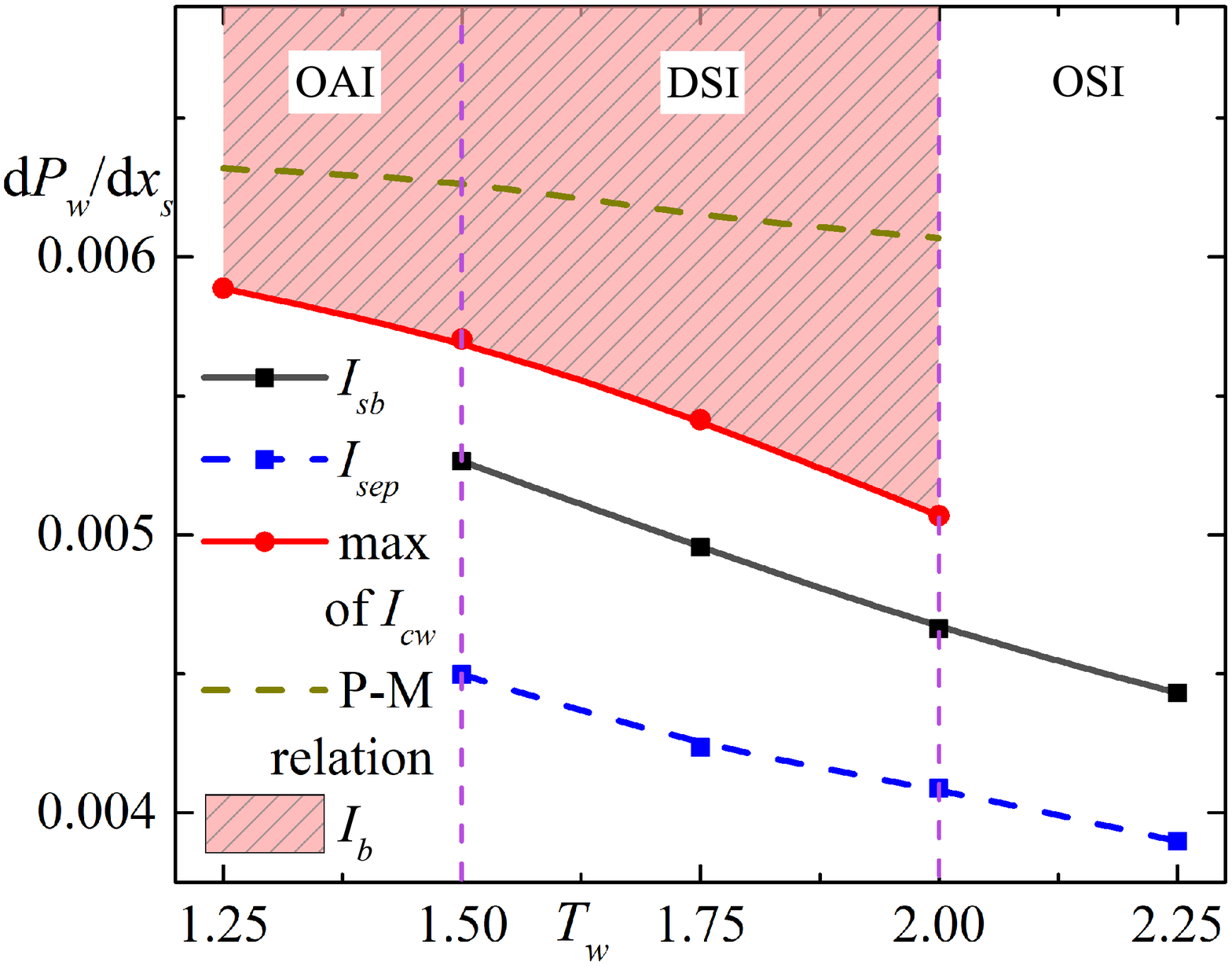}} \hspace{-2mm}
	\caption{characteristic pressure gradient: (a)$Ma_{\infty}$-variation, (b)$T_{w}$-variation. Black squares are $I_{sb}$ and red dots are the maximum of $I_{cw}$. The red shadow is the possible range of $I_{b}$ at the location of maximum of $I_{cw}$. Blue squares are the pressure gradients at separation points. The brown lines are the results of Eq.(\ref{eq:dPM}).} \label{fig:Ib}
\end{figure}

To confirm above theoretical analysis, we illustrate $I_{sb}(x_{sep})$, $I_{sb}(x_{max})$, the maximum of $I_{cw}$ and the range of $I_{b}$ of attachment states in Fig.\ref{subfig:Ib_Ma} and Fig.\ref{subfig:Ib_Tw}. In Fig.\ref{subfig:Ib_Ma}, it show that with $Ma_{\infty}$ increasing, $I_{sb}(x_{sep})<I_{sb}(x_{max})$ and both of them decrease, which are consistent with the above analysis. The red dots in Fig.\ref{subfig:Ib_Ma} are the maximum of  $I_{cw}$, which become smaller with $Ma_{\infty}$ decreasing. The APGs calculated from Eq.(\ref{eq:dPM}) at the location where the maximum of  $I_{cw}$ occurs in DNS are also provided and close to the DNS results. The deviation may due to the fact that the back propagation of the downstream boundary layer's weak APG under the sonic line has not been considered. With the Prop.\ref{pro:att}, $I_{cw} < I_{b}$ for the attachment states, thus the maximum of $I_{cw}$ bounded the corresponding local $I_{b}$.

With decreasing $T_{w}$, the increasing of $I_{sb}(x_{sep})$ and $I_{sb}(x_{max})$ shown in Fig.\ref{subfig:Ib_Tw} are consistent with the analysis. The maximum of $I_{cw}$ become smaller with increasing $T_{w}$, which is different from the constant value predition of Eq.(\ref{eq:dPM}).

Finally, we discuss three general characteristics in the hysteresis mechanism. The first characteristic is the non-universal trend of APG change during hysteresis. From the theoretical analysis and data validations, for the states pass from OSI to OAI, $I_{sb}$ grows with $Ma_{\infty}$ variation; but it drops with $T_{w}$ variation. This indicates that for different parameters variations, even if the corresponding processes are the same (flow state from OSI to OAI), the trend of APG change is not universal. The second characteristic is the incomparability of $I_{cw}$ and $I_{sb}$. The above discussion is based on the fact that both separation and attachment states are stable. The conditions of those stable states are related to the local relative magnitude of $I_{sb}$ and $I_{b}$ ($I_{cw}$ and $I_{b}$ for attachment states), and irrelevant to the relative magnitude of $I_{sb}$ and $I_{cw}$. Thus our analysis does not hypothesize $I_{cw} < I_{sb}$ in Ref.\cite{hu2020bistable}, and is consist with $I_{cw}$ and $I_{sb}$ distributions provided in Fig.\ref{fig:Ib}. The third characteristic is that the stable conditions of attachment and separation proposed by Prop.\ref{pro:att} and \ref{pro:sep} are independent of which parameter changes and how it changes. Thus the existence of separation hysteresis is guaranteed by the coexistence of Prop.\ref{pro:att} and \ref{pro:sep} in certain parameters interval and that $I_{sb}$, $I_{cw}$ and $I_{b}$ in the conditions have appropriate spatial distributions.



\subsection{Sonic Line Distributions}

\begin{figure}[htbp]
	\centering
	\subfigure[\label{subfig:sonicline_Ma}]{\includegraphics[width = 0.8\columnwidth]{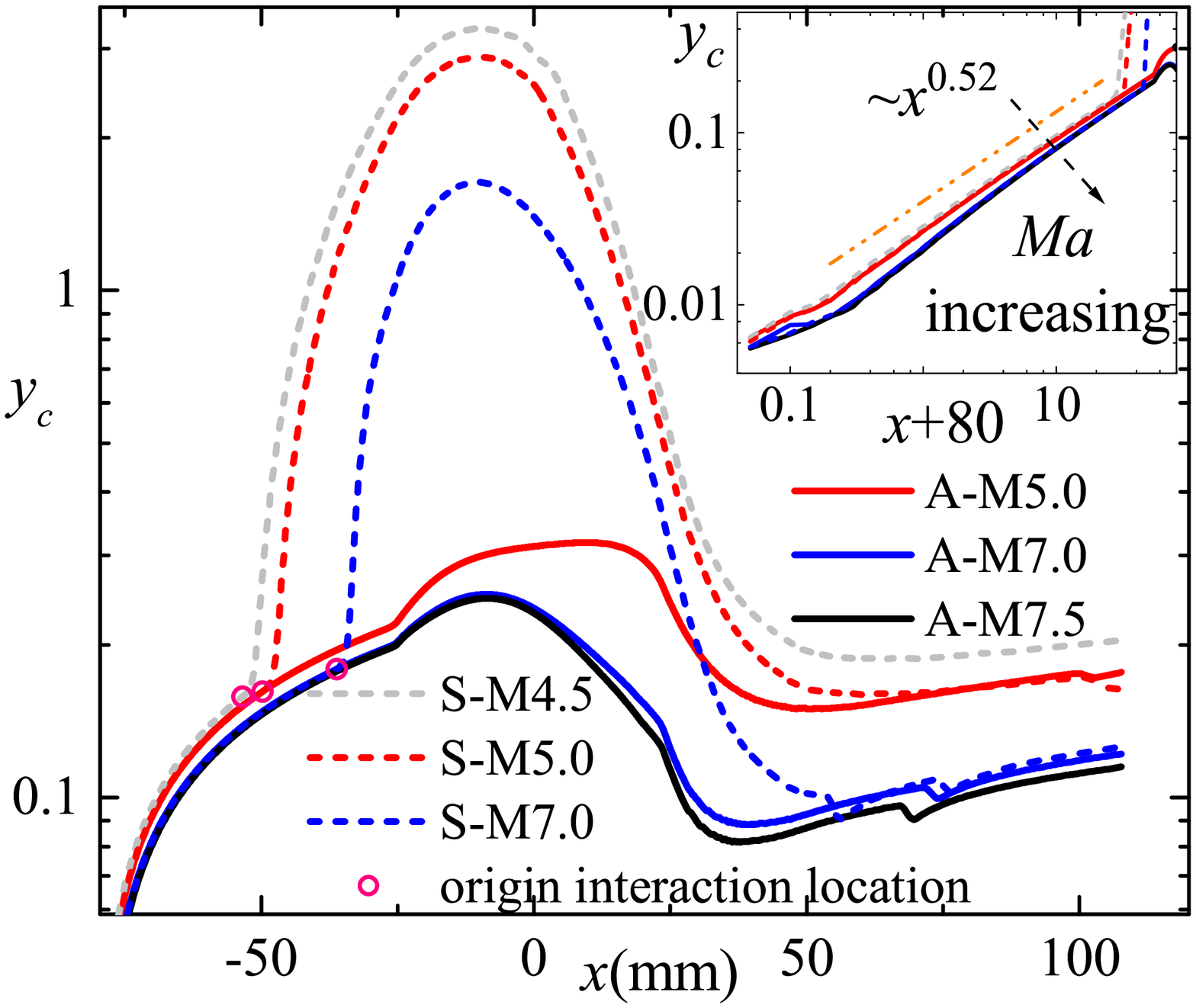}} 
	\hspace{4mm}
	\subfigure[\label{subfig:sonicline_Tw}]{\includegraphics[width = 0.8\columnwidth]{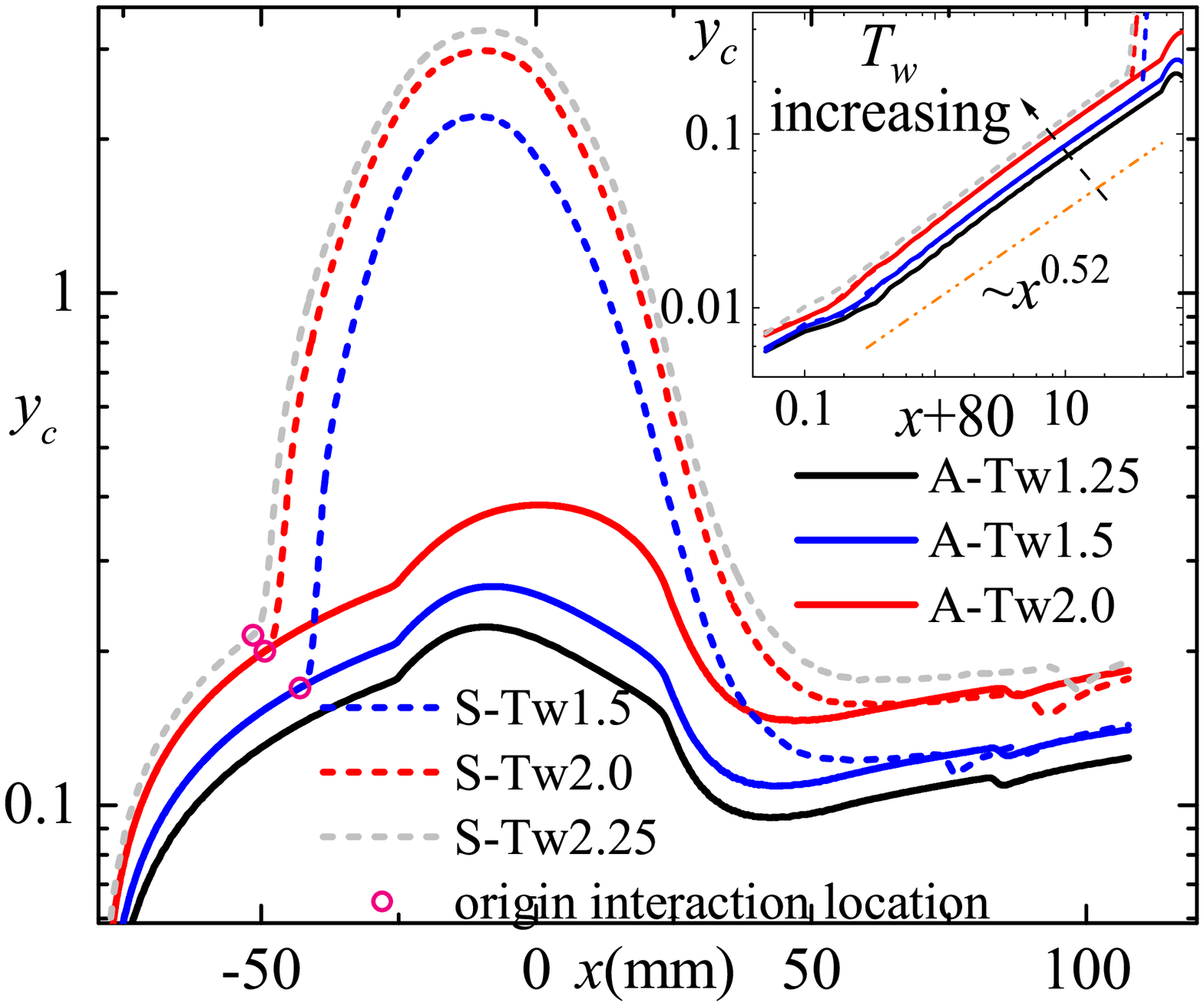}} \hspace{-2mm}
	\caption{The height of sonic lines $y_{c}$ for attachment and separation states. The inner figures are the zoom-in plots of the inflow boundary layer. The orange line indicates the scaling $(x+80)^{0.52}$. (a) $Ma_{\infty}$-variation; (b) $T_{w}$-variation.} \label{fig:SL}
\end{figure}

As analyzed above, the margin of OSI, DSI and OAI are determined by the relative magnitudes of $I_b(x)$ to $I_{sb}$ and $I_{cw}$. Here, $I_b(x)$ is deserved further consideration. The height of the sonic line $y_{c}$ is one of the indicators of the strength of boundary layer resistance $I_b(x)$, because it reflects the ``channel width'' that the downstream APG can propagate upstream. The higher the sonic line is, the farther the disturbance propagates upstream, which plays an important role in characterizing the stability of attachment and the influence of separation. In Fig.\ref{subfig:sonicline_Ma} and Fig.\ref{subfig:sonicline_Tw}, it can be seen that the upstream flow's sonic lines of separation and attachment states will coincide and increase slowly \cite{babinsky2011shock} for $x \leq x_{oi}$. Behind $x_{oi}$, separation states' sonic lines increase rapidly and reach the edge of the separation bubble till the flow reattaches to the wall, while the attachment states' sonic lines increase slowly and reach the maxima on the curved wall. The sonic lines of the separation states overlap with the attachment states again and increase slowly when the flow sweeps the tilt wall.

Generally, with $Ma_{\infty}$ increasing ($T_{w}$ decreasing), the sonic lines will be closer to the wall\cite{babinsky2011shock}. Because of the height of the sonic line $y_{c}(x)$ must satisfy $U(y_{c}(x))= a(y_{c}(x))$, where $U$ is the speed magnitude $U = \sqrt{u^2+v^2}$, $a=\sqrt{T}/Ma_{\infty}$ is the local acoustic velocity. As a rough approximation, the speed magnitude is proportional to the height away from the wall, the height of sonic line is inversely proportional to $Ma_{\infty}$ and proportional to $\sqrt{T_w}$. But in fact, with $Ma_{\infty}$ and $T_w$ increasing, the boundary layer becomes thicker, so the variation of sonic line is smaller than $Ma_{\infty}^{-1}$ but greater than $\sqrt{T_w}$. Moreover, based on the similarity solution of compressible Blasius Equation \cite{schlichting2000boundary}(with linear viscosity and $Pr \approx 1$ approximation), the streamwise variation of $y_{c}(x+80) \sim (x+80)^{1/2}$ can be derived. The DNS results consist with the analysis, shown in the inner figure of Fig.\ref{fig:SL}.

There is a qualitative relationship between the bistable phenomenon and the height of sonic line. For high $Ma_{\infty}$ or low $T_{w}$, the sonic lines lay close to the wall, thus the APG and disturbance  will be difficult to propagate upstream from the ramp to the weak resistance region (the curved wall, see $C_{f}$ in Fig.\ref{subfig:Cf_AS_Ma} and Fig.\ref{subfig:Cf_AS_Tw} and the shape factor discussed below) to separate the flow. When $Ma_{\infty}$ decreases to $Ma_{\infty, s}$ (or $T_{w}$ increases to $T_{w,s}$), the sonic line will rise high enough for the upstream propagating APG being larger than the boundary layer's resistance, then the flow will separate. 
For the separation states, because of the ``freezing characteristics''\cite{hu2020bistable} after the formation of the separation bubble, i.e., the separation bubble will not disappear suddenly if the separation is large enough. The downstream disturbance and APG have little effect on the separation, even if $Ma_{\infty}$ increases (or $T_w$ decreases) again. Thus, the height of sonic line affects the margin of DSI and OSI.

\subsection{Shape Factor Distributions}
As the compressible shape factor is strong dependent function of $Ma_{\infty}$\cite{babinsky2011shock,wenzel2019self}, the incompressible shape factor $H_{in}$ can be used to analyze the APG resistance of the bistable states\cite{babinsky2011shock}. It is defined as

\begin{equation} \label{eq:shapeFactor} 
H_{in}(x) = \frac{\delta_{in}^{*}(x)}{\delta_{in}^{**}(x)}
\end{equation}

\begin{figure}[htbp]
	\centering
	\subfigure[\label{subfig:Hfactor_Ma}]{\includegraphics[width = 0.8\columnwidth]{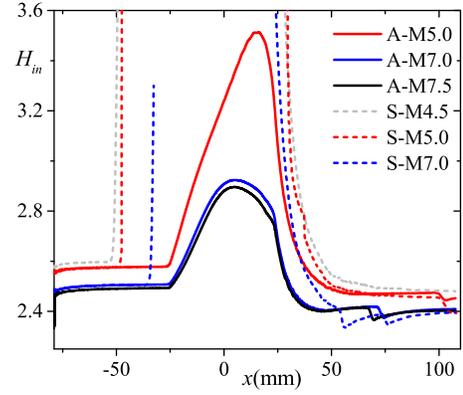}}
	\hspace{4mm}
	\subfigure[\label{subfig:Hfactor_Tw}]{\includegraphics[width = 0.8\columnwidth]{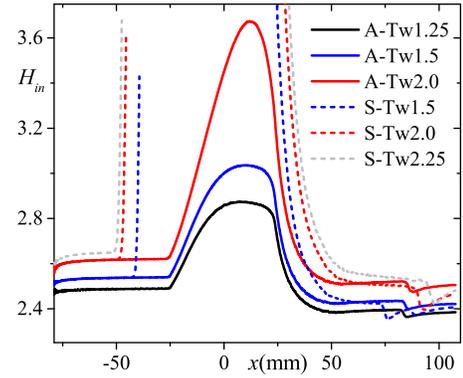}}
	\caption{Incompressible shape factor $H_{in}(x)$ for attachment and separation states. (a) $Ma_{\infty}$-variation; (b) $T_{w}$-variation.} 
\end{figure}

The distributions of $H_{in}(x)$ are shown in the Fig.\ref{subfig:Hfactor_Ma} and Fig.\ref{subfig:Hfactor_Tw}. For comparison, $H_{in}$ of incompressible laminar Blasius boundary layer is 2.59\cite{schlichting2000boundary}. The boundary layer's $H_{in}$ is approximately constant  before interaction. It increases from 2.49 to 2.59 when $Ma_{\infty}$ decreases from 7.5 to 4.5 and from 2.48 to 2.65 when $T_{w}$ increases from 1.25 to 2.25. The smaller $H_{in}(x)$ is, the fuller the boundary layer is, and the stronger its resistance to the APG is. Therefore, with $Ma_{\infty}$ increasing (or $T_w$ decreasing), the boundary layer is more likely to be attached. 

For the attachment states, $H_{in}(x)$ reaches its maximum on the curved wall, and converges to a smaller value (compared with the inflow $H_{in}$) for the ramp flow. For the boundary layer of the separation state, $H_{in}$ is the same as that of the attachment state for $x \leq x_{oi}$, but it rapidly diverges. After reattaching, it almost collapses with the attachment state's $H_{in}$ again.

Comparing Fig.\ref{subfig:Cf_AS_Ma}, \ref{subfig:sonicline_Ma} and \ref{subfig:Hfactor_Ma} for $Ma_{\infty}$-variation hysteresis (or Fig.\ref{subfig:Cf_AS_Tw}, \ref{subfig:sonicline_Tw} and \ref{subfig:Hfactor_Tw} for $T_{w}$-variation hysteresis), the synchro behaviors of $C_{f}$, $y_{c}(x)$ and $H_{in}(x)$ on the curved wall (not only the trend, but also the location of the critical positions) are all pointing to a self consistent result -- with increasing $Ma_{\infty}$ or decreasing $T_{w}$, the boundary layer APG resistance $I_{b}(x)$ is strengthened and eventually larger then $I_{cw}$ of OAI.

\section{Conclusion}
Separation hysteresis processes induced by inflow $Ma$ number $Ma_{\infty}$ and wall temperature $T_{w}$ variation are discovered in CCR flow and the hysteresis mechanism is analyzed in detail. When $Ma_{\infty,s} \leq Ma_{\infty} \leq Ma_{\infty,a}$ (or $T_{w,s} \geq T_{w} \geq T_{w,a}$), the boundary layer in CCR flow can be steadily attached or separated for the same IBCs, i.e., the state is determined by the initial condition. The corresponding parameter interval is called double solution interval (DSI). For the parameters outside this interval, only one stable state exists, either separation or attachment. Under the same conditions, the separation and attachment states lead to the great difference of shock configurations and aerothermodynamic features. If the attachment states can be maintained, the peak of wall heat flux can be reduced by 26\% at most within the present parameters.

The appearance of DSI is mainly determined by the relative magnitudes of APG induced by separation ($I_{sb}(x)$) and curved wall ($I_{cw}(x)$) to the boundary layer resistance ($I_{b}(x)$), which are changing with $Ma_{\infty}$ and $T_{w}$ variations. The $Ma_{\infty}$ and $T_{w}$ variations of $I_{sb}(x)$ and $I_{cw}(x)$ are analyzed via FIT and isentropic compression process, respectively, which are validated by the numerical simulations. With $Ma_{\infty}$ increasing, $I_{sb}(x)$ decreases and $I_{cw}(x)$ increases. $I_{b}(x)$ is characterized by the wall friction $C_{f}$ and incompressible shape factor $H_{in}(x)$ and affected by the height of sonic line $y_{c}(x)$. It is found that $I_{b}(x)$ drops with $Ma_{\infty}$ increasing or $T_{w}$ decreasing. The flow will leave DSI and go into OSI, if $I_{sb}(x), I_{cw}(x)$ and $I_{b}(x)$ do not fulfill Prop.\ref{pro:att}. The flow will go into OAI, if $I_{sb}(x), I_{cw}(x)$ and $I_{b}(x)$ do not fulfill Prop.\ref{pro:sep}.

In addition, the propositions of the stable conditions of separation and attachment (Prop.\ref{pro:att} and \ref{pro:sep}) are important to analyze the uniqueness and the non-uniqueness of separation states if $I_{b}(x)$ has been completely quantified in the future work. In fact, Multiple branches of separation have been found by triple deck theory\cite{braun2003Analysis} and simulation\cite{korolev1992Nonuniqueness}. It is expected that the multiple branches of separation will also occur in SBLI. The further analysis will provide a more comprehensive understanding of the separation hysteresis, which could be useful in wide speed range and long endurance flight.


\section*{Appendix: Definition of compressible/incompressible dispalcement thickness and momentum thickness}
The incompressible displacement thickness and momentum thickness are defined as 

\begin{equation} \label{eq:thick} 
\begin{aligned}
&\delta_{in}^{*}(x) = \int_{0}^{\delta} 1- \frac{u_{1}}{u_{1,e}}dy_{n}, \quad \\
&\delta_{in}^{**}(x) = \int_{0}^{\delta} \frac{u_{1}}{u_{1,e}} \left( 1- \frac{u_{1}}{u_{1,e}} \right)dy_{n}
\end{aligned}
\end{equation}

The compressible displacement thicknesses and momentum thicknesses are defined as 
\begin{equation} \label{eq:com_thick} 
\begin{aligned}
&\delta^{*}(x) = \int_{0}^{\delta} 1- \frac{\rho u_{1}}{\rho_{e} u_{1,e}}dy_{n}, \quad \\
&\delta_{in}^{**}(x) = \int_{0}^{\delta} \frac{\rho u_{1}}{\rho u_{1,e}} \left( 1- \frac{u_{1}}{u_{1,e}} \right)dy_{n}
\end{aligned}
\end{equation}
where $u_{1}=u\cos \varphi + v \sin \varphi$, $\varphi$ is defined in Eq.(\ref{eq:phi}) and $\varphi=0$ for flat plate, $\varphi=\phi$ for the tilt plate. And $\delta$ is the boundary layer thickness, whose edge is the position of nondimensional velocity gradient less than $10^{-2}$ in the velocity profile\cite{wenzel2019self,spalart2000mechanisms}. Because the CCR flow involves the deflection of the velocity profile by the external shock wave, it is inaccurate to use the traditional $\delta_{99}$ (where velocity reaches 99\% of the free stream velocity) to define the boundary layer thickness. The threshold choice of $10^{-2}$ is empirical that it has little effect on the magnitudes of displacement and momentum thickness. $u_{1,e}$ and $\rho_{e}$ are the velocity and density at the edge of boundary layer, respectively.

\begin{acknowledgments}
We are grateful to professor Xin-Liang Li and You-Sheng Zhang for their helpful discussions. This work was supported by the National Key R \& D Program of China (Grant No. 2019YFA0405300). Yan-Chao Hu thanks to the support of the China Postdoctoral Science Foundation (Grant No. 2020M683746).
\end{acknowledgments}

\section*{Data availability}
The data that support the findings of this study are available from the corresponding author upon reasonable request.

\bibliography{MaTwHyster}

\end{document}